\documentclass[preprint2]{aastex6}

\usepackage{natbib}
\begin{document}

\title{The Peculiar Multi-Wavelength Evolution Of V1535 Sco}
\author{J.~D.~Linford\altaffilmark{1,2}, L.~Chomiuk\altaffilmark{3}, T.~Nelson\altaffilmark{4}, T. Finzell\altaffilmark{3}, F.~M. Walter\altaffilmark{5}, J.~L.~Sokoloski\altaffilmark{6}, K.~Mukai\altaffilmark{7,8}, A.~J.~Mioduszewski\altaffilmark{9}, A.~J. van der Horst\altaffilmark{1,2}, J.~H.~S.~Weston\altaffilmark{11}, and M.~P.~Rupen\altaffilmark{10}}

\altaffiltext{1}{Department of Physics, The George Washington University, Washington, DC 20052, USA}
\altaffiltext{2}{Astronomy, Physics, and Statistics Institute of Sciences, The George Washington University, Washington, DC 20052, USA}
\altaffiltext{3}{Department of Physics and Astronomy, Michigan State University, East Lansing, MI 48824, USA}
\altaffiltext{4}{Department of Physics and Astronomy, University of Pittsburgh, Pittsburgh, PA 15260}
\altaffiltext{5}{Department of Physics and Astronomy, Stony Brook University, Stony Brook, NY 11794, USA}
\altaffiltext{6}{Columbia Astrophysics Laboratory, Columbia University, New York, NY, USA}
\altaffiltext{7}{Center for Space Science and Technology, University of Maryland Baltimore County, Baltimore, MD 21250, USA}
\altaffiltext{8}{CRESST and X-ray Astrophysics Laboratory, NASA/GSFC, Greenbelt MD 20771 USA}
\altaffiltext{9}{National Radio Astronomy Observatory, P.O. Box 0, Socorro, NM 87801, USA}
\altaffiltext{10}{Green Bank Observatory, P.O. Box 2, Green Bank, WV 24944, USA}
\altaffiltext{11}{Herzberg Institute of Astrophysics, National Research Council of Canada, Penticton, BC, Canada}
\email{jlinford@.gwu.edu}

\begin{abstract}
We present multi-wavelength observations of the unusual nova V1535 Sco throughout its outburst in 2015.  Early radio observations were consistent with synchrotron emission, and early X-ray observations revealed the presence of high-energy (\textgreater 1 keV) photons.  These indicated that strong shocks were present during the first $\sim$2 weeks of the nova's evolution.  The radio spectral energy distribution was consistent with thermal emission from week 2 to week 6.  Starting in week 7, the radio emission again showed evidence of synchrotron emission and there was an increase in X-ray emission, indicating a second shock event.  The optical spectra show evidence for at least two separate outflows, with the faster outflow possibly having a bipolar morphology.  The optical and near infrared light curves and the X-ray N$_{\rm H}$ measurements indicated that the companion star is likely a K giant.

\end{abstract}
\keywords{white dwarfs --- novae, cataclysmic variables --- stars: individual (V1535 Sco) --- radio continuum: stars --- X-rays: stars}

\section{Introduction}
\label{intro}

A nova eruption occurs when a white dwarf has accreted enough material from a companion star to trigger thermonuclear runaway on its surface (e.g., Starrfield et al. 1972).  
Observations with modern telescopes have revealed the presence of strong shocks in nova systems.  In particular, the discovery of GeV $\gamma$-rays from V407 Cyg indicated that in at least some systems, the shocks were strong enough to accelerate particles to very high energies (Abdo et al. 2010).  In the case of V407 Cyg, the shocks were the result of the nova ejecta colliding with a dense stellar wind from a Mira variable companion (Abdo et al. 2010).  Much to the surprise of the nova community, further observations with the \emph{Fermi Gamma-ray Space Telescope} discovered $\gamma$-ray emission from systems with main sequence companions (e.g., Ackermann et al. 2014).

While observations across the electromagnetic spectrum have contributed to the discovery of shocks in novae, radio observations have proven to be particularly valuable.  Resolved radio images can reveal non-spherical structure and provide evidence for multiple outflows, as in the case of V959 Mon (Chomiuk et al. 2014).  Early monitoring of the radio emission, even when the ejecta are too small to resolve, has also prove to be an effective probe of shocks in novae.  The shocks can manifest as synchrotron emission with non-thermal brightness temperatures in spectacular ways as in V1723 Aql (Weston et al. 2016a), or more subtle ways as in V5589 Sgr (Weston et al. 2016b).  When the radio synchrotron emission is bright enough, the shocks can be directly imaged with very long baseline interferometry, as in V959 Mon (Chomiuk et al. 2014) and RS Oph (Rupen et al. 2008; Sokoloski et al. 2008).

Observations with modern X-ray telescopes have also proven to be useful in characterizing shocks in novae.  Those novae embedded within the wind of a giant companion typically emitting bright, hard X-rays early in their evolution, such as RS Oph (Bode et al. 2006; Nelson et al. 2008) and V745 Sco (Page et al. 2015; Orio et al. 2015).  Novae in systems with main sequence companions are also capable of producing strong X-rays, but typically later in their evolution Such was the case in both V382 Vel (Mukai \& Ishida 2001) and V5589 Sgr (Weston et al. 2016b).

On 2015 February 11, the nova V1535 Sco was discovered in the constellation Scorpius.  It was reported in vsnet-alert 18276 by P. Schmeer and in CBET 4078 by T. Kojina.  Early optical and near infrared follow-up indicated the nova was fading rapidly (Walter 2015).  The combination of rapid fading and a possible bright near-infrared counterpart indicated that the companion star could be an M giant, in which case the nova could be embedded in the wind from a giant companion (Walter 2015).  Srivastava et al. (2015) later argued that the companion could be a K giant based on pre-eruption 2MASS photometry of a likely counterpart (2MASS J17032617-3504178).   Spectroscopic observations indicated it was a He/N type nova (Walter 2015; Srivastava et al. 2015).  Early X-ray and radio observations (Nelson et al. 2015) indicated the presence of strong shocks in the ejecta, which was further evidence that the ejecta were expanding into a dense medium.  The nova was originally given the designation PNV J1703260-3504140, then known as Nova Sco 2015, and eventually given the official designation V1535 Sco.  Srivastava et al. (2015) also noted that narrowing of the H-line profiles in the infrared indicated the presence of a deceleration shock, and estimated the total mass ejected to be between $4.5\times10^{-6}$ and $2.6\times10^{-4}$ $\epsilon$ M$_{\odot}$ (where $\epsilon$ is the filling factor of the ejecta).

During the 2015 outburst, no \textit{Fermi Gamma-ray Space Telescope} staring mode observations were scheduled and the nova was not reported as a detection from the survey mode data.  However, it must be noted that \textit{Fermi} is generally less sensitive to transients at the Galactic Center due to a combination of the high background, and the north-south ``rocking'' profile of the survey mode results in the Galactic Center having minimal exposure (Acero et al. 2015).  It is therefore possible that V1535 Sco produced $\gamma$-rays like other novae (e.g., V407 Cyg and V745 Sco; Abdo et al. 2010; Cheung et al. 2014), but none were detectable due to these limiting factors.  Recent work by Morris et al. (2017) argues that all novae produce $\gamma$-ray emission, but we mainly detect the nearby ones with \emph{Fermi}.

In this paper, we present our multi-wavelength observations of V1535 Sco made during 2015.  These observations were made with the Karl J. Jansky Very Large Array (VLA), the Very Long Baseline Array (VLBA), the \textit{Swift} X-ray Telescope (XRT), and the Small \& Moderate Aperture Research Telescope System (SMARTS).  
We present our observations and data reduction methods in Section 2.  Our knowledge about the distance to V1535 Sco is discussed in Section 3.  In Section 4, we present results from the multi-wavelength observations.  We discuss our findings in Section 5.  Final conclusions are summarized in Section 6.  Throughout this paper, we use the initial optical detection of the nova on 2015 February 11.837 (MJD 57064.837) as Day 0.0.  We also use the term ``embedded nova'' to refer to any nova embedded in the wind material from its companion star (e.g. Chomiuk et al. 2012; Mukai et al. 2014).

\section{Observations and Data Reduction}
\subsection{VLA Observations}
We began monitoring V1535 Sco with the VLA within 64 hours of its discovery in the optical.  The first epoch only included observations at C-band (4.0 - 8.0 GHz) and Ka-band (26.5 - 40.0 GHz).  All following epochs also included observations at L-band (1.0-2.0 GHz) and Ku-band (12.0 - 18.0 GHz).  To maximize spectral coverage, all bands were split into upper and lower sidebands.  For the first 8 epochs, we used 8-bit sampling for all frequencies.  This gave us a total bandwidth of 1.0 GHz at L-band, and 2.0 GHz at all other bands.  For the ninth epoch (Day 93.463) we used 3-bit sampling for Ku-band and Ka-band, resulting in total bandwidths of 6.0 GHz and 8.0 GHz, respectively.  For the final epoch (Day 122.363), we used 3-bit sampling for C-band, Ku-band, and Ka-band, giving total bandwidths of 4.0 GHz, 6.0 GHz, and 8.0 GHz, respectively.  10 epochs were observed under the program VLA/13B-057, and one epoch (Day 93.463) was observed under the program S61420.  The nova was detected during the first epoch (Day 2.7) and remained detectable at multiple frequencies for nearly 100 days.  We ceased observations once the nova had faded to the point where it was no longer detected at all frequencies.
\begin{center}
\begin{deluxetable*}{ccccccc}
\tablewidth{0 pt}
\tabletypesize{\footnotesize}
\setlength{\tabcolsep}{0.025in}
\tablecaption{ \label{radiotab}
VLA Observations}
\tablehead{UT Date & Day\tablenotemark{1} & VLA & Central Frequency & Time on Source & Flux Density & Uncertainty \\
&  & Config & (GHz) & (minutes) & (mJy) & (mJy) }
\startdata
2015-02-14 & 2.663 & B & 4.55 & 13.6 & 4.13 & 0.13 \\
 & & & 7.38 & 13.6 & 2.786 & 0.085 \\
 & & & 28.2 & 11.3 & 0.819 & 0.071 \\
 & & & 36.5 & 11.3 & 0.675 & 0.090 \\
2015-02-18 & 6.663 & B & 13.5 & 11.9 & 0.416 & 0.021 \\
 & & & 17.4 & 11.9 & 0.344 & 0.024 \\
 & & & 28.2 & 11.8 & 0.295 & 0.053 \\
 & & & 36.5 & 11.8 & 0.376 & 0.074 \\
2015-02-19 & 7.663 & B & 1.26 & 15.6 & 1.57 & 0.090 \\
 & & & 1.74 & 15.6 & 1.21 & 0.081 \\
 & & & 4.55 & 13.6 & 0.650 & 0.026 \\
 & & & 7.38 & 13.6 & 0.439 & 0.018 \\
2015-02-24 & 12.763 & B & 13.5 & 12.5 & 0.385 & 0.020 \\
 & & & 17.4 & 12.5 & 0.456 & 0.024 \\
 & & & 28.2 & 12.4 & 0.785 & 0.062 \\
 & & & 36.5 & 12.4 & 0.845 & 0.087 \\
2015-03-01 & 17.663 & B & 1.26 & 15.6 & \textless0.317 & 0.10 \\
 & & & 1.74 & 15.6 & \textless0.317 & 0.066 \\
 & & & 4.55 & 13.6 & 0.221 & 0.019 \\
 & & & 7.38 & 13.6 & 0.192 & 0.017 \\
2015-03-07 & 23.663 & B & 13.5 & 12.0 & 0.585 & 0.024 \\
 & & & 17.4 & 12.0 & 0.755 & 0.031 \\
 & & & 28.2 & 11.9 & 1.15 & 0.074 \\
 & & & 36.5 & 11.9 & 1.68 & 0.11 \\
2015-03-10 & 26.563 & B & 1.26 & 15.6 & \textless0.374 & 0.074 \\
 & & & 1.74 & 15.6 & \textless0.271 & 0.054 \\
 & & & 4.55 & 13.5 & 0.455 & 0.023 \\
 & & & 7.38 & 13.5 & 0.536 & 0.021 \\
2015-03-25 & 41.583 & B & 13.5 & 12.0 & 0.267 & 0.015 \\
 & & & 17.4 & 12.0 & 0.299 & 0.017 \\
 & & & 28.2 & 12.0 & 0.348 & 0.044 \\
 & & & 36.5 & 12.0 & \textless0.351 & 0.055 \\
2015-03-25 & 41.633 & B & 1.26 & 15.6 & \textless0.609 & 0.112 \\
 & & & 1.74 & 15.6 & \textless0.271 & 0.060 \\
 & & & 4.55 & 13.8 & 0.222 & 0.017 \\
 & & & 7.38 & 13.8 & 0.253 & 0.014 \\
2015-04-07 & 54.563 & B & 13.5 & 12.5 & 0.457 & 0.028 \\
 & & & 17.4 & 12.5 & 0.399 & 0.022 \\
 & & & 28.2 & 12.4 & 0.323 & 0.047 \\
 & & & 36.5 & 12.4 & 0.276 & 0.062 \\
2015-04-08 & 55.733 & B & 1.26 & 15.7 & 0.87 & 0.10 \\
 & & & 1.74 & 15.7 & 0.661 & 0.060 \\
 & & & 4.55 & 13.7 & 0.483 & 0.024 \\
 & & & 7.38 & 13.7 & 0.374 & 0.018 \\
2015-04-18 & 65.513 & B & 1.26 & 16.6 & 0.444 & 0.082 \\
 & & & 1.74 & 16.6 & 0.258 & 0.057 \\
 & & & 4.55 & 14.6 & 0.256 & 0.019 \\
 & & & 7.38 & 14.6 & 0.228 & 0.014 \\
2015-04-19 & 66.493 & B & 13.5 & 12.0 & 0.209 & 0.017 \\
 & & & 17.4 & 12.0 & 0.167 & 0.023 \\
 & & & 28.2 & 12.0 & 0.173 & 0.049 \\
 & & & 36.5 & 12.0 & \textless0.289 & 0.083 \\
2015-05-01 & 78.463 & B & 1.26 & 15.6 & \textless0.32 & 0.11 \\
 & & & 1.74 & 15.6 & \textless0.378 & 0.071 \\
 & & & 4.55 & 13.6 & 0.170 & 0.019 \\
 & & & 7.38 & 13.6 & 0.139 & 0.015 \\
2015-05-01 & 78.503 & B & 13.5 & 12.1 & 0.129 & 0.017 \\
 & & & 17.4 & 12.1 & 0.164 & 0.022 \\
 & & & 28.2 & 12.0 & 0.174 & 0.053 \\
 & & & 36.5 & 12.0 & \textless0.241 & 0.075 \\
2015-05-16 & 93.463 & B$\rightarrow$BnA & 1.26 & 8.0 & \textless0.531 & 0.177 \\
 & & & 1.74 & 8.0 & \textless0.336 & 0.078 \\
 & & & 4.55 & 8.0 & 0.0727 & 0.023 \\
 & & & 7.38 & 8.0 & 0.0708 & 0.017 \\
 & & & 13.5 & 8.0 & 0.0518 & 0.018 \\
 & & & 16.5 & 8.0 & \textless0.0859 & 0.021 \\
 & & & 29.5 & 9.0 & \textless0.149 & 0.036 \\
 & & & 35.0 & 9.0 & \textless0.135 & 0.045 \\
2015-06-14 & 122.363 & BnA$\rightarrow$A & 1.26 & 8.0 & \textless0.78 & 0.12 \\
 & & & 1.74 & 8.0 & \textless0.233 & 0.078 \\
 & & & 5.0 & 8.0 & \textless0.0806 & 0.020 \\
 & & & 7.0 & 8.0 & \textless0.111 & 0.022 \\
 & & & 13.5 & 8.0 & \textless0.106 & 0.016 \\
 & & & 16.5 & 8.0 & \textless0.120 & 0.021 \\
 & & & 29.5 & 9.0 & \textless0.264 & 0.045 \\
 & & & 35.0 & 9.0 & \textless0.176 & 0.059 
\enddata
\tablenotetext{1}{We take the time of initial detection 2015 February 11.837 UT (MJD 57064.837) to be Day 0.0}
\end{deluxetable*}
\end{center}

For all epochs, we used the absolute flux calibrator 3C286.  Complex gains calibrators were J1626-2951 for both L-band and C-band, and J1650-2943 for both Ku-band and Ka-band.  All VLA data were calibrated with the NRAO VLA calibration pipeline, version 1.3.1, which uses CASA\footnote{http://www.casa.nrao.edu} version 4.2.2.  The pipeline script was executed on either a dedicated desktop or an NRAO Lustre node.  Once calibrated, the data were exported to AIPS\footnote{http://www.aips.nrao.edu} for additional flagging.  The fully-flagged data were imaged with Difmap (Sheperd 1997).  The images were then imported back into AIPS and flux densities were measured using the task \verb|JMFIT|.  The L-band and C-band data from 2015 April 18 were also calibrated in AIPS in order to look for polarization.  Also, the Ku-band and Ka-band data from 2015 March 25 were re-calibrated in AIPS to check the results of the CASA pipeline.  Our VLA results are presented in Table~\ref{radiotab}.  For non-detections, we calculated the upper limit as the flux density at the nova location plus 3 times the image root mean square (rms).  Our uncertainties are calculated by adding the image rms and an absolute flux density uncertainty in quadrature.  Perley \& Butler (2013) report that the VLA absolute flux density calibration is stable to within $1\%$ for 1 to 20 GHz, and within $3\%$ for 20 to 50 GHz.  Because V1535 Sco is far from 3C286, our uncertainty in the flux density calibration will be higher.  We adopt an absolute flux density uncertainty of $3\%$ for 1 to 20 GHz, and $5\%$ for above 20 GHz.
\begin{center}
\begin{deluxetable*}{cccccc}
\tablewidth{0 pt}
\tabletypesize{\footnotesize}
\setlength{\tabcolsep}{0.025in}
\tablecaption{ \label{vlbatab}
VLBA Observations}
\tablehead{UT Date & Day\tablenotemark{1} & Central Frequency & Time on Source & Flux Density & Uncertainty \\
&  & (GHz) & (minutes) & (mJy) & (mJy) }
\startdata
2015-02-19 & 7.663 & 4.87 & 177.3 & 0.477 & 0.056  \\
2015-02-24 & 12.763 & 4.87 & 177.3 & \textless0.278 & 0.041 \\
\enddata
\tablenotetext{1}{We take the time of initial detection 2015 February 11.837 UT (MJD 57064.837) to be Day 0.0}
\end{deluxetable*}
\end{center}
\begin{center}
\begin{deluxetable*}{ccccccc}
\tablewidth{0 pt}
\tabletypesize{\footnotesize}
\setlength{\tabcolsep}{0.025in}
\tablecaption{ \label{swiftab1}
\emph{Swift}/XRT Observations}
\tablehead{Obs ID & Day\tablenotemark{1} & Exposure & Total Counts & 0.3-1.0 keV & 1-10 keV & Total \\
&  & (s) &  & Count Rate & Count Rate & Count Rate }
\startdata
00033634002 & 4.16 & 4084 & 591 & 0.0062 & 0.1578 & 0.1640 \\
00033634003 & 11.04 & 3485 & 198 & 0.0555 & 0.0878 & 0.1428 \\
00033634004 & 13.94 & 3407 & 431 & 0.0886 & 0.0553 & 0.1429 \\
00033634005 & 17.93 & 4669 & 676 & 0.1342 & 0.0329 & 0.1666 \\
00033634006 & 24.75 & 2959 & 252 & 0.0925 & 0.0174 & 0.1091 \\
00033634007 & 31.51 & 4378 & 46 & 0.0102 & 0.0067 & 0.0161 \\
00033634008 & 38.77 & 4568 & 27 & 0.0021 & 0.0054 & 0.0075 \\
00033634009 & 45.55 & 2138 & \nodata & \textless0.0011 & \textless0.0011 & \textless0.0082 \\
00033634010 & 49.08 & 2575 & 5 & 0.0008 & 0.0021 & 0.0024 \\
00033634011 & 53.07 & 4714 & 12 & 0.0016 & 0.0025 & 0.0041 \\
00033634012 & 60.03 & 4845 & 13 & 0.0013 & 0.0015 & 0.0027 \\
00033634013 & 66.71 & 2979 & 7 & 0.0007 & 0.0026 & 0.0028 \\
00033634014 & 73.86 & 4934 & \nodata & \textless0.0005 & \textless0.0002 & \textless0.0036 \\
00033634015 & 80.32 & 4526 & \nodata & \textless0.0002 & \textless0.0001 & \textless0.0024 \\
\enddata
\tablenotetext{1}{We take the time of initial detection 2015 February 11.837 UT (MJD 57064.837) to be Day 0.0}
\end{deluxetable*}
\end{center}

\subsection{VLBA Observations}
We had 2 epochs of VLBA observations under the program VLBA/15A-269, both in C-band with a central frequency of 4.87 GHz and a total bandwidth of 256 MHz.  For both epochs, we used J1709-3525 as the phase reference source.  The bright sources J1656-3302 and J1713-3418 were also observed as a means of gauging the successful calibration of the nova.  For the first epoch, 3 antennas were not usable: Hancock, North Liberty, and Mauna Kea.  For the second epoch, Hancock and Mauna Kea could again not be used.  The total time on source for each epoch was approximately 177.3 minutes (or 2.96 hours).

Both epochs were calibrated using standard routines in AIPS, including the new bandpass correction routine described by Walker (2014).  Images were made in Difmap.  The nova was detected in the first epoch, but not in the second epoch (see Table~\ref{vlbatab}).  Schedules for further epochs were submitted, but conflict with a top-priority VLBA program prevented them from being observed.  The position of the compact source detected in the first VLBA epoch was RA 17$^{\text{h}}$03$^{\text{m}}$26$^{\text{s}}$.17218 $\pm$0$^{\text{s}}$.00002, DEC -35\arcdeg04\arcmin17.87267\arcsec $\pm$0.00071\arcsec.

\subsection{Swift Observations}
We obtained 14 epochs of \emph{Swift} XRT observations.  The exposure times ranged from 2.5 to 4.9 ks.  The nova was detected by the XRT in 11 of the 14 observations.  Details of the observations are given in Table~\ref{swiftab1}.

\begin{figure}
\includegraphics[width=3.3in]{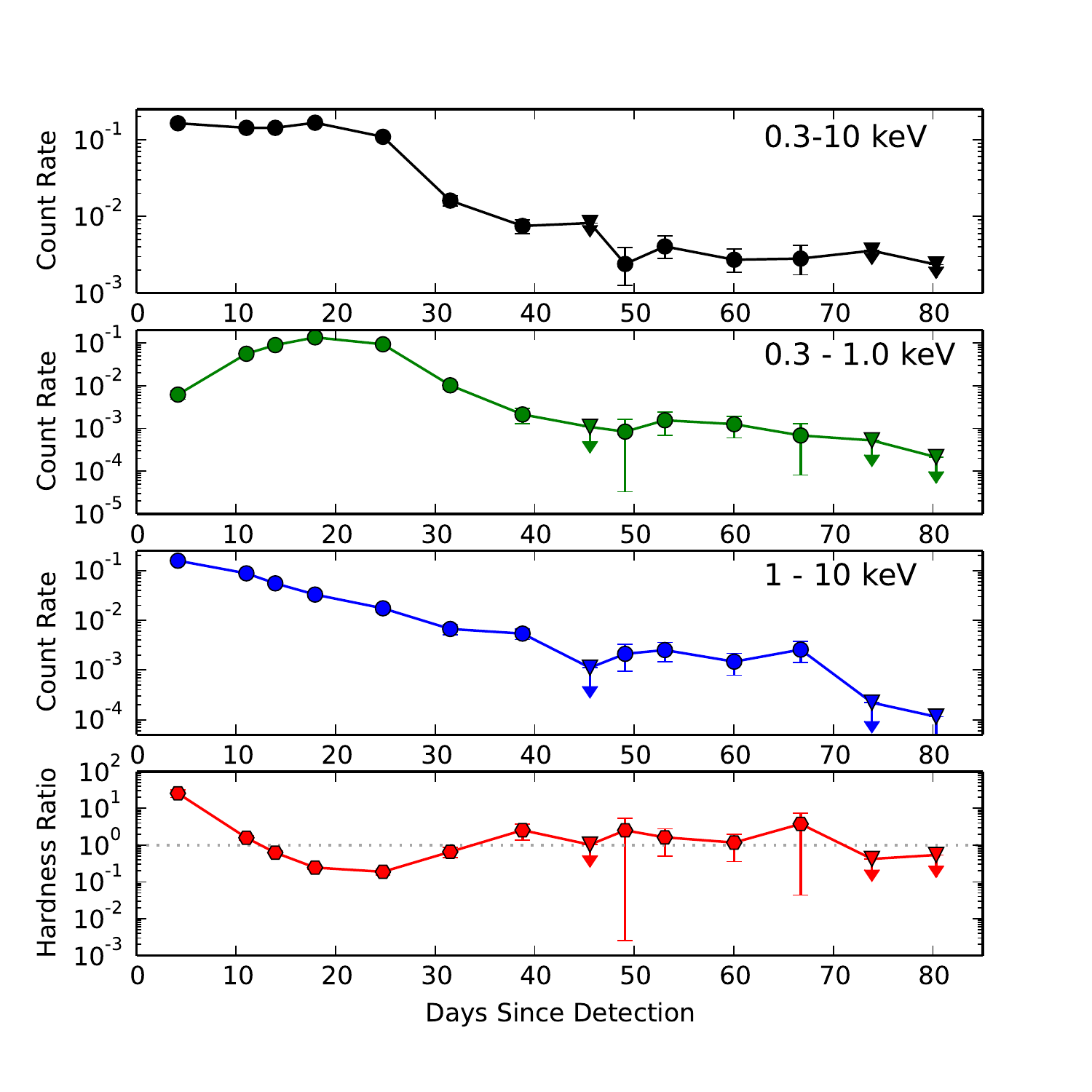}
\caption{Top panel: Total count rate from \emph{Swift} XRT. Second panel: Count rate in the 0.3-1.0 keV energy range. Third panel: Count rate in the 1.0-10.0 keV energy range.  Bottom panel: Hardness ratio (1.0-10 keV / 0.3-1.0 keV).  The dotted line indicates a ratio of 1.0.}
\label{swift_all}
\end{figure}

We divided the \emph{Swift} XRT detections into soft and hard bands based on the photon energy, $E$. We chose $E$\textless1.0 keV to be soft and $E$\textgreater1.0 keV to be hard.  Our decision to use these designations is based on previous observations of novae which indicate which indicate that the $E$\textless1.0 keV band can be dominated by the super-soft X-rays, while the $E$\textgreater1.0 keV X-rays are likely to be from shocks (e.g., Mukai et al. 2008; Mukai et al. 2014).  The X-ray emission was initially hard.  The soft emission increased over the first $\sim$20 days while the hard emission decreased over the same time period.  We had no detections with the XRT after Day 67.  Our \emph{Swift} XRT results are plotted in Figure~\ref{swift_all}.

\begin{figure}
\includegraphics[trim=10 10 10 10,clip,width=3.5in]{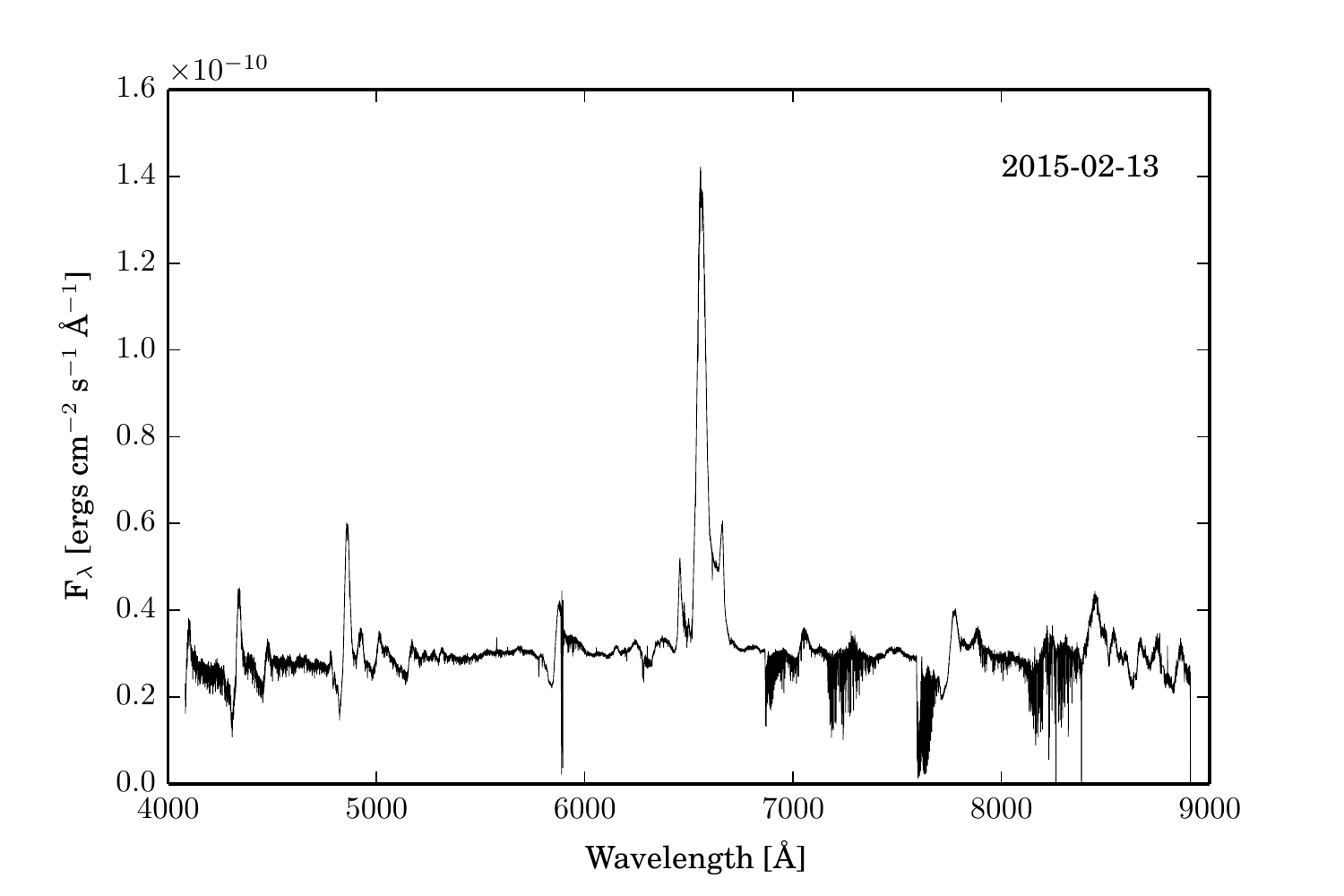}
\caption{Example optical spectrum of V1535 Sco obtained with the SMARTS telescope.  This spectrum was from Day 1.565.  Note the high velocity outliers on H$\alpha$ (6563 \AA).}
\label{smarts_exspec}
\end{figure}

\begin{figure}
\includegraphics[trim=0 10 10 20,clip,width=3.2in]{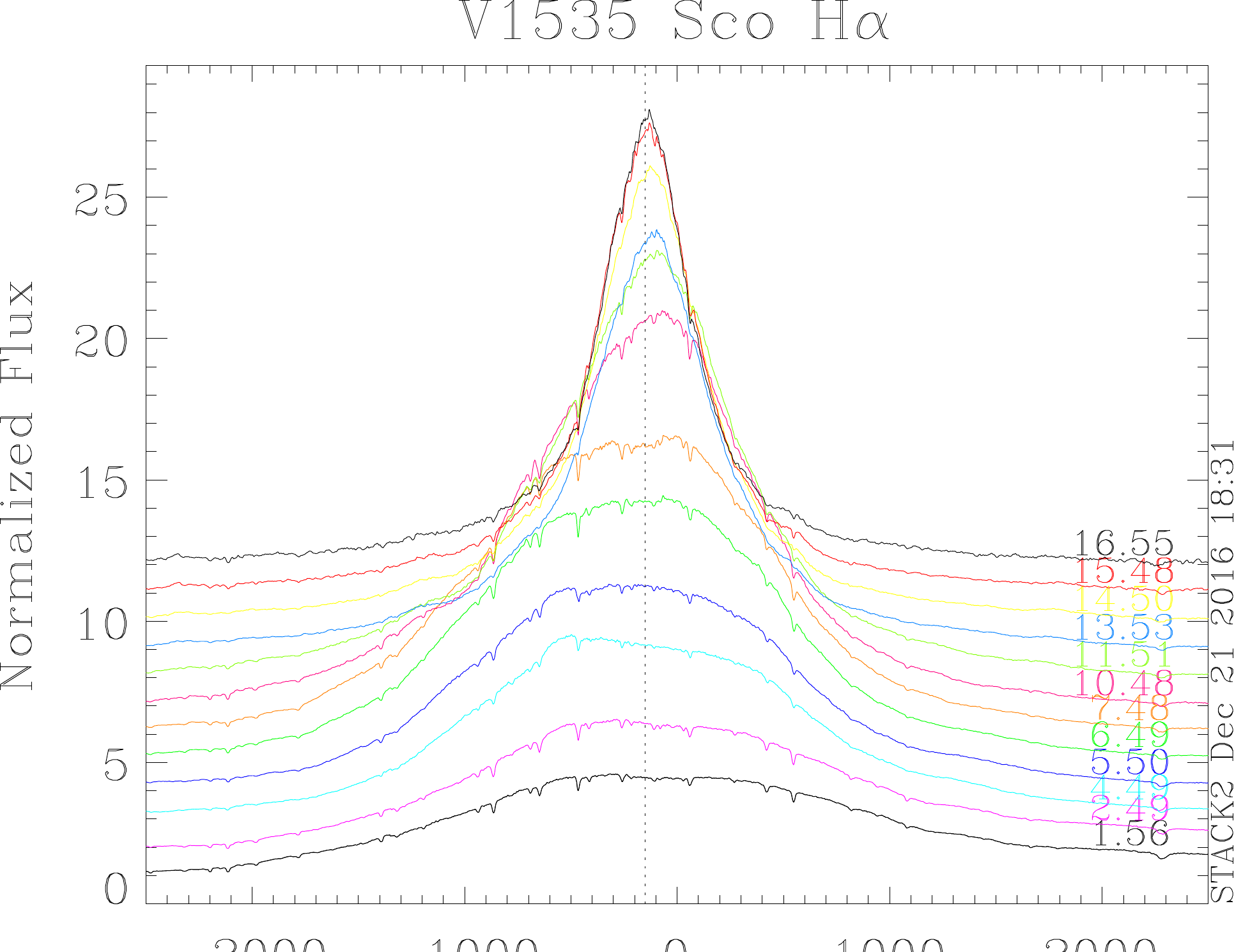}
\caption{The H$\alpha$ line over the first 16.5 days of nova eruption.  Note the rapid narrowing of the line as the fastest material becomes optically thin.}
\label{smarts_ha}
\end{figure}

\subsection{SMARTS Photometric and Spectroscopic Observations}
SMARTS monitors novae in the optical and near infrared and provides the results to the public via the Stony Brook/SMARTS Spectroscopic Atlas of (mostly) Southern Novae\footnote{www.astro.sunysb.edu/fwalter/SMARTS/NovaAtlas/atlas.html}. Photometric observations of V1535 Sco began on 2015 February 13.35 using the ANDICAM\footnote{http://www.astronomy.ohio-state.edu/ANDICAM/detectors.html} dual-channel imager on the SMARTS 1.3m telescope.

Spectroscopic measurements of the ejecta began on 2015 February 13.5 as part of the SMARTS observation campaign using the CHIRON echelle spectrometer on the 1.5m SMARTS telescope (Tokovinin et al. 2013).  As an example, we show the spectrum from Day 1.565 in Figure~\ref{smarts_exspec}.  There is P Cyg absorption present on the H$\beta$ (4861 \AA), H$\gamma$ (4340 \AA), \ion{He}{1} (4471, 4921,\& 5879 \AA), \ion{Fe}{2} (5169 \AA), and \ion{O}{1} (7774 \& 8446 \AA) lines.  This is unusual for a He/N nova, as they typically do not have much mass loss.  

\begin{figure}
\includegraphics[width=3.5in]{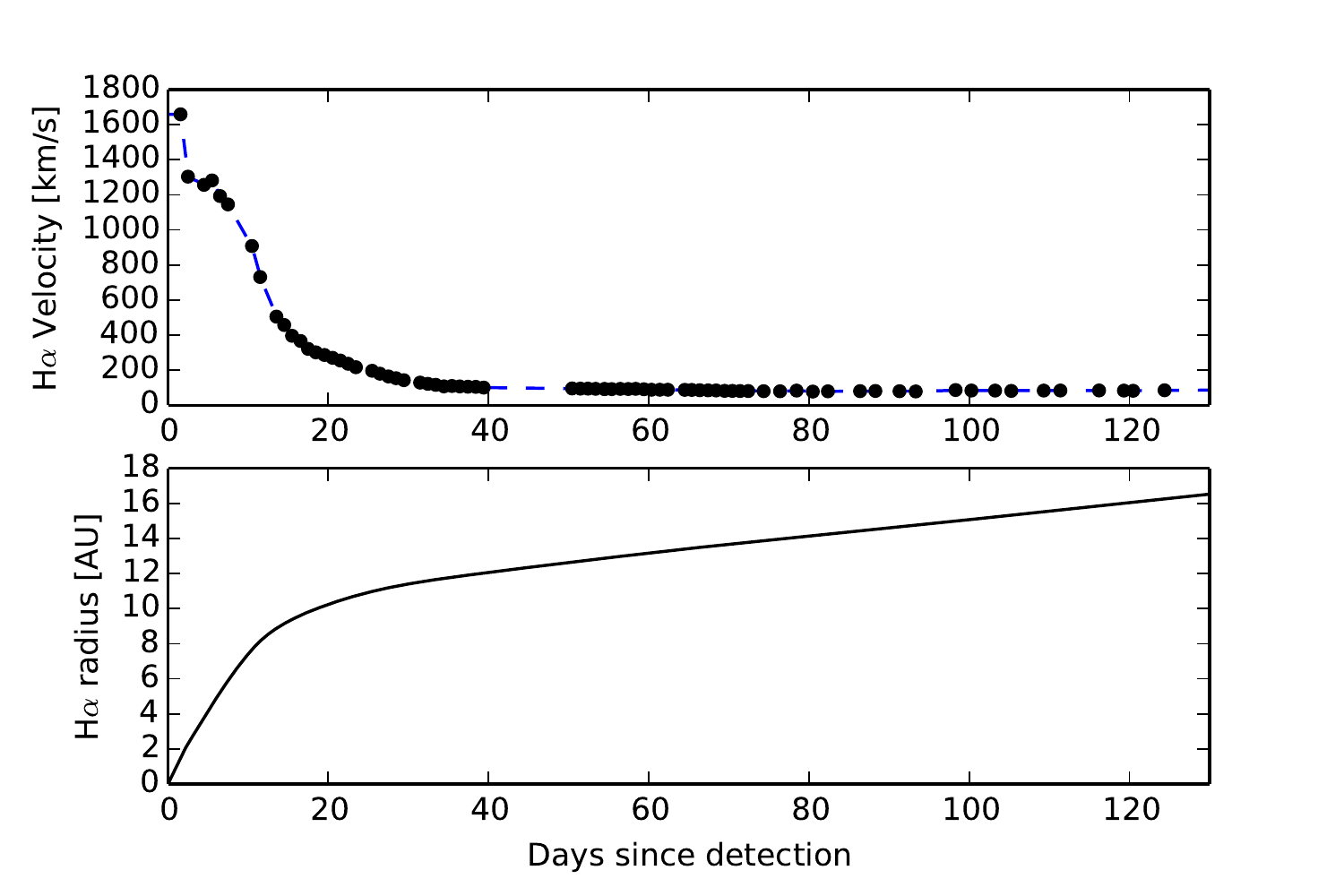}
\caption{{\bf Upper panel:} Velocity profile from the H$\alpha$ line width of V1535 Sco.  The dashed blue line shows the interpolation of the data used to estimate the radius of the H$\alpha$ emitting region. {\bf Bottom panel:} Integrated radial size of the H$\alpha$ emitting region.}
\label{sco15_vnr}
\end{figure}

The H$\alpha$ line width measurements are presented in Table~\ref{smarts_hatab}, and we show the evolution of the H$\alpha$ line from Day 1.565 to Day 16.548 in Figure~\ref{smarts_ha}.  Because we are concerned with the outermost edge of the main component of the ejecta, we estimate the maximum velocity as the full width at 3$\sigma$ (FW3$\sigma$).  We convert the H$\alpha$ FWHM velocities to FW3$\sigma$ velocities by assuming the core of the spectral line has a Gaussian shape, then we multiply by a factor of $1.5(2 \ln 2)^{-1/2}$ in order to get the $3\sigma$ velocities.  Our velocity evolution is shown in Figure~\ref{sco15_vnr}.  While the measured velocity of the ejecta appears to decrease with time, we must emphasize that what is actually being measured here is the velocity of the H$\alpha$ emitting region, which does not necessarily correspond to the outer edge of the ejecta.  

\begin{figure}
\includegraphics[trim=0 10 10 25,clip,width=3.2in]{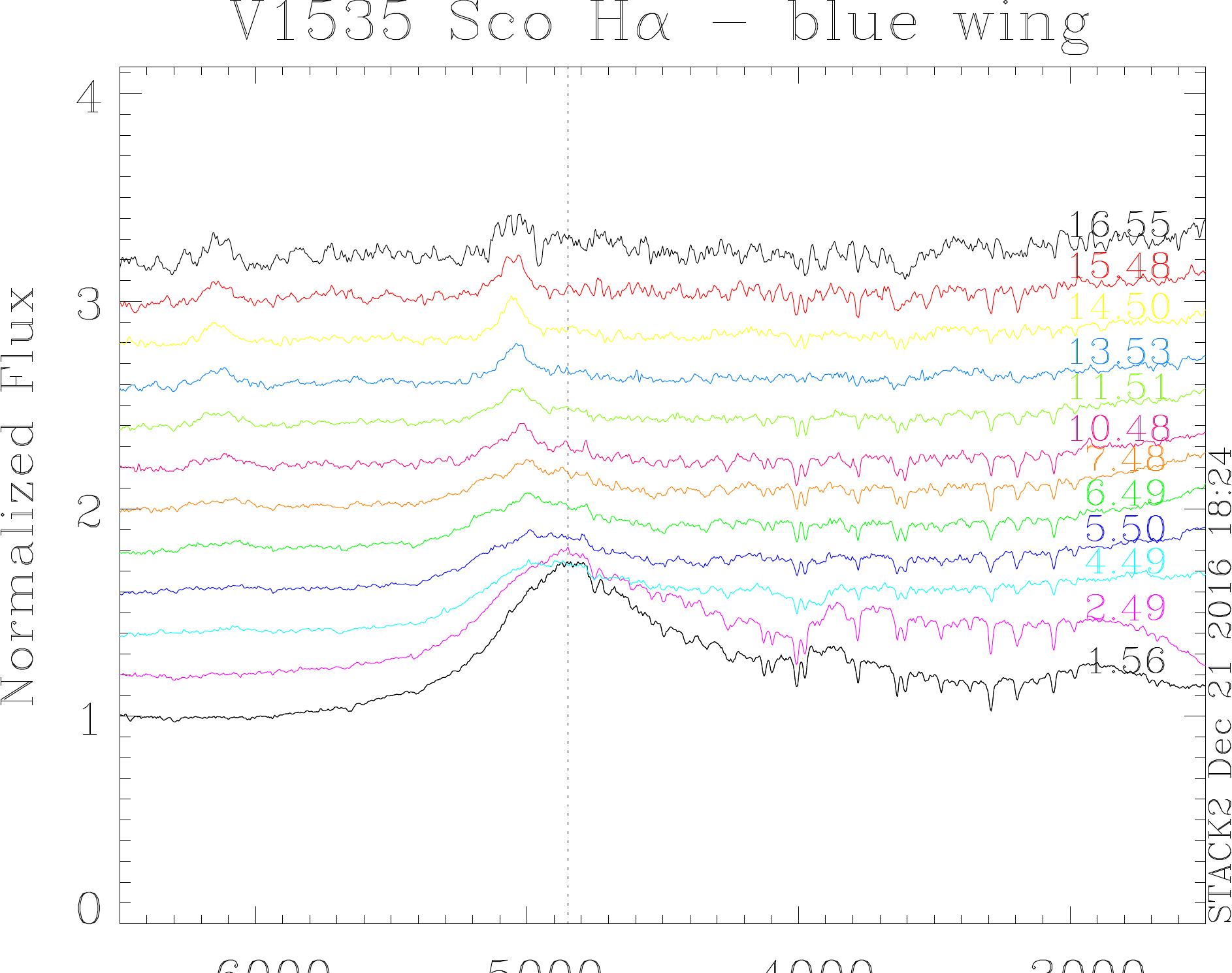}
\caption{The evolution of the blue-shifted high velocity H$\alpha$ outlier over the first 16.5 days of the nova eruption.  The vertical line is at -4850 km s$^{-1}$.  Note that by Day 6.5 the \ion{Fe}{2} lines at 6455.8 and 6456.4 \AA\, dominate this region of the spectrum.}
\label{smarts_hab}
\end{figure}

\begin{figure}
\includegraphics[trim=0 10 10 25,clip,width=3.2in]{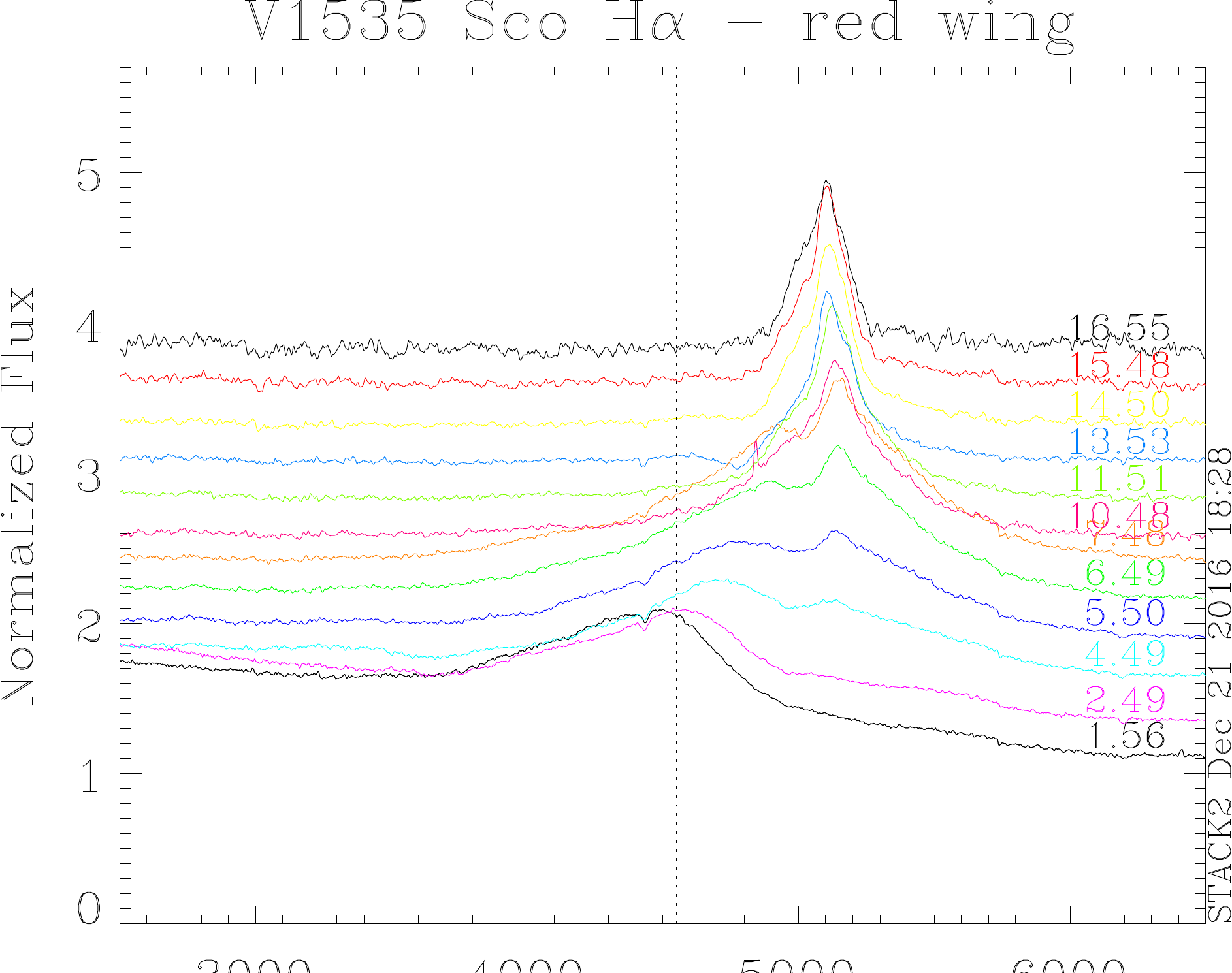}
\caption{The evolution of the red-shifted high velocity H$\alpha$ outlier over the first 16.5 days of the nova eruption.  The vertical line is at 4550 km s$^{-1}$.  Note that the peak of the outlier clearly moves to higher velocities through Day 6.5.  Also, note that the \ion{He}{1} line at 6678 \AA\, appears on Day 4.5 and dominates this region of the spectrum by Day 7.5.}
\label{smarts_har}
\end{figure}

The H$\alpha$ (6563 \AA) line shows high velocity outliers (both red and blue shifted; see Table~\ref{smarts_hawings}).  These outliers do not appear to decelerate as long as they are present.  In fact, the high velocity outliers appear to accelerate throughout the first week of the nova's evolution, although this could be the result of blending with lines from \ion{He}{1} (6678 \AA) and \ion{Fe}{2} (6455.8 \& 6456.4 \AA) which are not detectable until later in the nova's evolution.  We show the evolution of the high velocity outliers in Figures~\ref{smarts_hab} (blue-shifted) and \ref{smarts_har} (red-shifted).  These outliers could be explained by a shell of high velocity material.  However, the narrowness of the outlier lines would require that the shell had a very small thickness, and such a fast-moving thin shell should become diffuse (and thus undetectable) much faster than we see here.  Another explanation is a bipolar outflow, where the narrowness of these high velocity outliers would point to a relatively well-collimated outflow.  It would also likely have a fairly small inclination angle for us to detect such large velocities.  Such well-collimated bipolar outflows have been seen in several other novae.  Examples of embedded novae with bipolar outflows are U Sco (L\'{e}pine et al. 1999) and RS Oph (Rupen et al. 2008; Sokoloski et al. 2008).  Non-embedded novae are also known to have bipolar outflows, such as V1494 Aql (Iijima \& Esenoglu 2003), V475 Sct (Kawabata et al. 2006), V445 Pup (Woudt et al. 2009), V5668 Sgr (Bannerjee et al. 2016), and possibly T Pyx (Chesneau et al. 2011). 

The high velocity outliers in Figures~\ref{smarts_exspec}, \ref{smarts_hab}, and \ref{smarts_har} are likely to be H$\alpha$.  We see marginal evidence for similar outliers with identical velocities around the H$\beta$ line.  Were these the [\ion{N}{2}] lines, the 6482 \AA\, line would be blueshifted by approximately 1150 km s$^{-1}$, while the 6668 \AA\, line would be redshifted by only about 300 km s$^{-1}$.  Low velocity \ion{Fe}{2} and \ion{He}{1} lines appear  in these regions around day 5 and strengthen as the ejecta evolve.  Lacking other plausible identifications, we identify the outliers as H$\alpha$ ejected at high velocities.

Using the H$\alpha$ line width as the approximate velocity evolution for the optically-thick component of the ejecta, we can use $r(t_{1})=\int_{0}^{t_{1}}v(t)dt$ to determine the approximate radial size of the ejecta at any given time $t_{1}$.  We perform this integration numerically using the trapezoidal approximation.  First, we interpolate our velocity measurements to give a smoother function to integrate over.  We assume that the velocity is constant between Day 0 and our first spectroscopic measurement.  This is likely an oversimplification, although it may not be drastically far from the average velocity over this time, as the ejecta must first be accelerated to some (unknown) maximum velocity, then decelerate to the velocity we measured with our first spectroscopic observation.  Our resulting radial size of the H$\alpha$ emitting region is shown in the bottom panel of Figure~\ref{sco15_vnr}.

The radial size of the H$\alpha$ emitting region can serve as a minimum size of the ejecta.  A more realistic approximation of the radial size of the ejecta is to assume that the ejecta expand at roughly the velocity from the first observation: $\sim$1659 km s$^{-1}$.  However, neither of these simple models account for the high velocity outliers seen in the early spectra.

\begin{deluxetable}{ccccc}
\tablewidth{0 pt}
\tabletypesize{\footnotesize}
\setlength{\tabcolsep}{0.025in}
\tablecaption{ \label{smarts_hatab}
SMARTS H$\alpha$ Spectral Measurements}
\tablehead{Day\tablenotemark{1} & FWHM & FWHM Velocity & FW3$\sigma$ & FW3$\sigma$ Velocity \\
& (\AA)  & (km s$^{-1}$) & (\AA) & (km s$^{-1}$) }
\startdata
1.565 & 47.5 & 1302 & 60.5 & 1659 \\
2.491 & 37.3 & 1023 & 47.5 & 1304 \\
4.493 & 36.0 & 987 & 45.8 & 1257 \\
5.495 & 36.7 & 1006 & 46.7 & 1282 \\
6.495 & 34.1 & 936 & 43.5 & 1193 \\
7.484 & 32.8 & 899 & 41.8 & 1145 \\
10.478 & 26.0 & 713 & 33.1 & 908 \\
11.505 & 20.9 & 574 & 26.7 & 731 \\
13.528 & 14.5 & 397 & 18.4 & 505 \\
14.505 & 13.1 & 359 & 16.7 & 458 \\
15.483 & 11.3 & 311 & 14.4 & 397 \\
16.548 & 10.5 & 288 & 13.4 & 366 \\
17.482 & 9.2 & 252 & 11.7 & 322 \\
18.456 & 8.6 & 237 & 11.0 & 302 \\
19.525 & 8.2 & 225 & 10.5 & 287 \\
20.53 & 7.7 & 212 & 9.9 & 271 \\
21.498 & 7.3 & 200 & 9.3 & 255 \\
22.472 & 6.8 & 186 & 8.6 & 237 \\
23.451 & 6.2 & 170 & 7.9 & 217 \\
25.477 & 5.6 & 154 & 7.2 & 196 \\
26.445 & 5.2 & 141 & 6.6 & 180 \\
27.507 & 4.7 & 129 & 6.0 & 164 \\
28.462 & 4.4 & 121 & 5.6 & 154 \\
29.435 & 4.1 & 112 & 5.2 & 143 \\
31.469 & 3.7 & 101 & 4.7 & 129 \\
32.432 & 3.5 & 96 & 4.5 & 122 \\
33.416 & 3.3 & 91 & 4.2 & 116 \\
34.432 & 3.1 & 85 & 3.9 & 108 \\
35.44 & 3.1 & 86 & 4.0 & 110 \\
36.42 & 3.1 & 84 & 3.9 & 107 \\
37.432 & 3.0 & 83 & 3.9 & 106 \\
38.413 & 3.0 & 82 & 3.8 & 105 \\
39.398 & 2.9 & 79 & 3.7 & 101 \\
50.435 & 2.7 & 75 & 3.5 & 96 \\
51.468 & 2.7 & 74 & 3.5 & 95 \\
52.412 & 2.7 & 74 & 3.5 & 95 \\
53.364 & 2.7 & 74 & 3.4 & 94 \\
54.463 & 2.7 & 73 & 3.4 & 93 \\
54.505 & 2.6 & 72 & 3.4 & 92 \\
55.389 & 2.6 & 72 & 3.4 & 92 \\
56.447 & 2.7 & 73 & 3.4 & 93 \\
57.438 & 2.7 & 73 & 3.4 & 93 \\
58.371 & 2.7 & 74 & 3.4 & 94 \\
59.398 & 2.6 & 72 & 3.3 & 91 \\
60.343 & 2.5 & 69 & 3.2 & 88 \\
61.382 & 2.5 & 69 & 3.2 & 88 \\
62.381 & 2.5 & 69 & 3.2 & 88 \\
64.495 & 2.5 & 69 & 3.2 & 88 \\
65.371 & 2.5 & 69 & 3.2 & 87 \\
66.37 & 2.5 & 67 & 3.1 & 86 \\
67.413 & 2.4 & 67 & 3.1 & 85 \\
68.407 & 2.4 & 66 & 3.1 & 84 \\
69.473 & 2.4 & 64 & 3.0 & 82 \\
70.434 & 2.3 & 64 & 3.0 & 82 \\
71.4 & 2.3 & 64 & 3.0 & 81 \\
72.422 & 2.3 & 64 & 3.0 & 81 \\
74.341 & 2.3 & 63 & 2.9 & 80 \\
76.393 & 2.3 & 63 & 2.9 & 80 \\
78.453 & 2.4 & 66 & 3.0 & 84 \\
80.477 & 2.2 & 61 & 2.8 & 78 \\
82.36 & 2.3 & 62 & 2.9 & 79 \\
86.377 & 2.3 & 63 & 2.9 & 80 \\
88.287 & 2.3 & 64 & 3.0 & 81 \\
91.302 & 2.3 & 63 & 2.9 & 80 \\
93.322 & 2.3 & 62 & 2.9 & 79 \\
98.299 & 2.5 & 68 & 3.2 & 87 \\
100.274 & 2.4 & 66 & 3.1 & 84 \\
103.232 & 2.4 & 66 & 3.0 & 84 \\
105.25 & 2.3 & 64 & 3.0 & 82 \\
109.298 & 2.4 & 66 & 3.0 & 84 \\
111.376 & 2.4 & 66 & 3.1 & 84 \\
116.206 & 2.4 & 66 & 3.1 & 84 \\
119.308 & 2.4 & 66 & 3.0 & 84 \\
120.457 & 2.4 & 65 & 3.0 & 83 \\
\enddata
\tablenotetext{1}{We take the time of initial detection 2015 February \\11.837 UT (MJD 57064.837) to be Day 0.0}
\end{deluxetable}

\begin{deluxetable}{ccccc}
\tablewidth{0 pt}
\tabletypesize{\footnotesize}
\setlength{\tabcolsep}{0.025in}
\tablecaption{ \label{smarts_hawings}
SMARTS H$\alpha$ Shoulders}
\tablehead{Day\tablenotemark{1} & $\lambda_{blue}$ & Blue-shifted Velocity & $\lambda_{red}$ & Red-shifted Velocity \\
& (\AA)  & (km s$^{-1}$) & (\AA) & (km s$^{-1}$) }
\startdata
1.565 & 6458.25 & 4778 & 6658.67 & 4377 \\
2.491 & 6458.09 & 4785 & 6660.00 & 4414 \\
4.493 & 6456.03 & 4880 & 6665.90 & 4707 \\
5.495 & 6455.45 & 4906 & 6668.08 & 4807 \\
6.495 & 6455.5 & 4904 & 6669.98 & 4893 \\
7.484 & 6456.3 & 4867 & 6670.75 & 4929 \\
10.478 & 6456.52 & 4857 & 6671.3 & 4954 \\
11.505 & 6456.91 & 4839 & \nodata & \nodata \\
\enddata
\tablenotetext{1}{We take the time of initial detection 2015 February \\11.837 UT (MJD 57064.837) to be Day 0.0}
\end{deluxetable}


\section{The Distance to V1535 Sco}
The distance to V1535 Sco is highly uncertain.  Srivastava et al. (2015) apply the maximum magnitude rate of decline (MMRD) relations from della Valle \& Livio (1995) and Downes \& Duerbeck (2000) to get distance estimates of $13.7\pm0.4$ and $14.7\pm3.8$ kpc, respectively.  
However, Munari et al. (2017) reported a distance to V1535 Sco of approximately 9.7 kpc, despite using the same MMRD method as Srivastava et al. (2015).  This indicates the MMRD distance may be highly dependent on the data used, as Munari et al. (2017) performed all their observations on a single telescope and Srivastava et al. (2015) obtained their data from multiple telescopes via the American Association of Variable Star Observers (AAVSO).  Furthermore, substantial uncertainty has been shed on the MMRD method in general by the work of Kasliwal et al. (2011) and Cao et al. (2012).  The Srivastava et al. MMRD distances seem unlikely as they place the nova on the opposite side of the Galactic Bulge, which should make it extremely reddened and obscured due to the intervening material.  

We were able to constrain the distance to V1535 Sco by using pre-outburst photometry to determine a spectral type for the companion. The source 2MASS1703261-350417 is within $\leq 0.1$" of V1535 Sco, and has $J=13.423 \pm 0.037$, $H = 12.500 \pm 0.033$, and $K=12.190 \pm 0.041$. Without correcting for reddening, we find near-IR colors of $J-K = 1.233 \pm 0.078$ and $H-K = 0.310 \pm 0.074$. 

Using the derived reddening from Srivastava et al. (2015) of $E(B-V) = 0.72 \pm 0.05$ we can compute the necessary color corrections. The reddening conversion functions of Schlafly \& Finkbeiner (2011) give $E(J-K) = 0.413\times E(B-V)$, $E(H-K) = 0.15\times E(B-V)$ and $A_J = 0.723\times E(B-V)$. Using these, we find the dereddened colors to be $(J-K)' = 0.936 \pm 0.098$, $(H-K)' = 0.202 \pm 0.081$, and $J' = 12.902 \pm 0.073$, where the prime signifies that it is dereddened. Using the spectral class system of Covey et al.(2007), the colors are consistent with a spectral class from K3 III to M0 III which, in turn, give an absolute $J$ magnitude between $-3.92$ to $-4.75$. This gives a distance modulus of $m-M = 15.212 \pm 1.253$, or a distance between 6.2 and 19.6 kpc.  If we instead use $E(B-V) = 0.96$, as reported in Munari et al. (2017), our distance range is 5.8 to 14.0 kpc.

Note that, although we cannot definitively rule out the companion being a main sequence star using this method, if it were a main sequence progenitor system the distance modulus would be $<8$ mags. Since the peak brightness of the nova was $m_V \approx 9.5$, the absolute magnitude at peak would be $M_V > 1.0$, which is far too dim for a nova. Therefore, we can reasonably rule out a main sequence companion.  We therefore find it more likely that the companion star is an evolved star in the range between subgiant and giant. 

Because the progenitor distance is consistent with being located at the Galactic Center, and that is also where we expect most novae to occur, we will assume a distance of $\sim8.5$ kpc for V1535 Sco.  This distance agrees well with estimates in Munari et al. (2017) based on both MMRD and the Buscombe \& de Vaucouleurs (1955) method of estimating the absolute magnitude 15 days after optical maximum.

\section{Results}
\subsection{VLA Light Curve and Spectral Indices}
\begin{figure}
\includegraphics[width=3.5in]{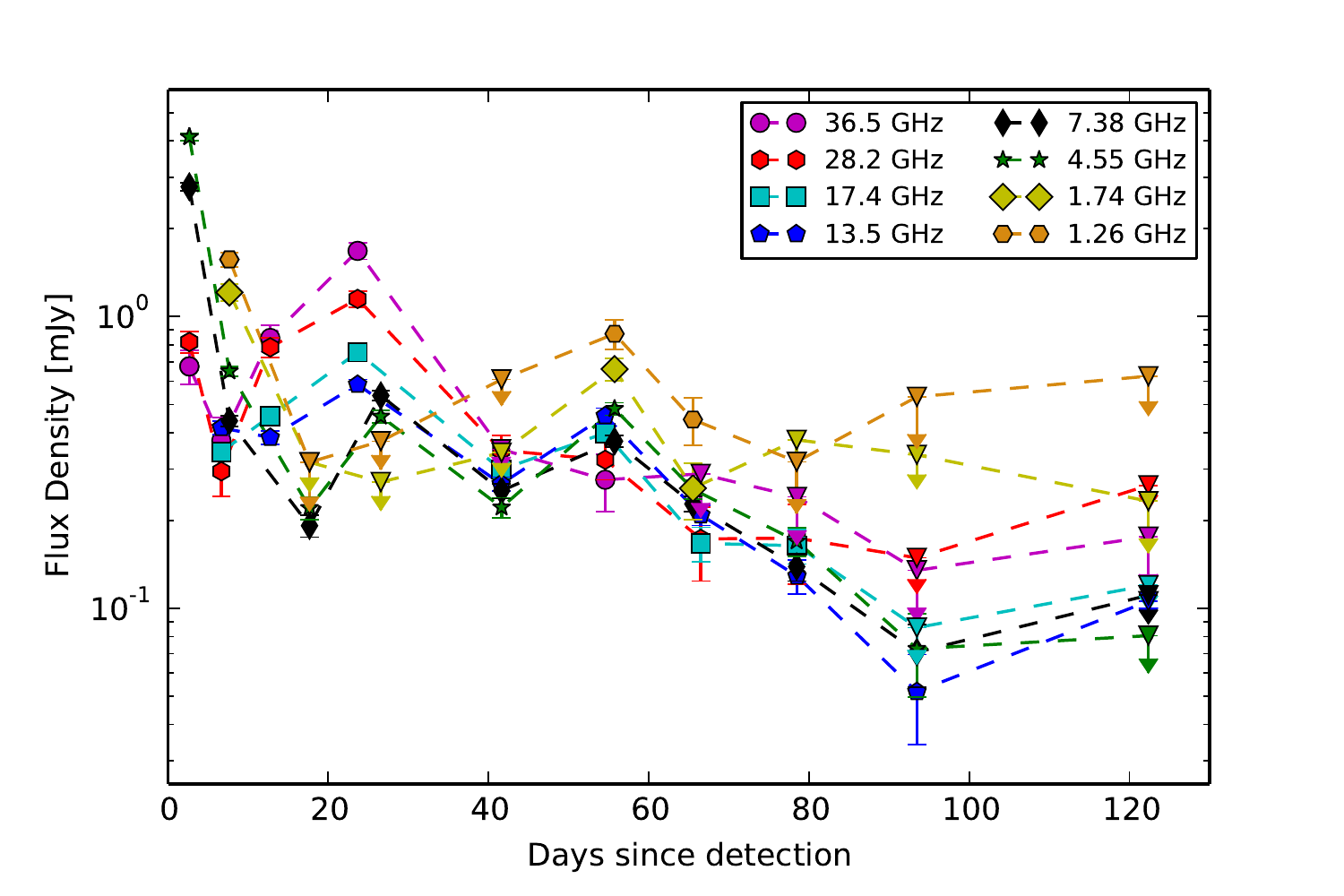}
\caption{Full light curves for the VLA observations.  Initial detection was 2015 February 11.837 UT (MJD 57064.837), which is used as $t=0$.  Non-detections are indicated with a downward-pointing triangle.}
\label{vlalc}
\end{figure}

The full VLA light curves for V1535 Sco are presented in Figure~\ref{vlalc}.  The nova was initially detected at relatively high flux densities, faded substantially, and then rose again to secondary peak around Day 25.  This secondary peak is unusual for novae, but not unprecedented (e.g., Taylor et al. 1987; Krauss et al. 2011; Eyres et al. 2009; Weston et al. 2016; Finzell et al. 2017).  After Day 27, the source faded again until about Day 56 when there is a tertiary peak.  This is highly unusual.  After the tertiary peak, the nova fades until it is no longer detectable.

The radio light curve for V1535 Sco is significantly different from the radio light curve of a non-embedded nova.  In a ``typical'' non-embedded nova, the radio emission is delayed from the optical emission by up to $\sim$2 weeks, and then slowly rises to a maximum over several months.  This is due to the fact that non-embedded novae are typically \emph{thermal} sources at radio wavelengths.  Therefore, their flux density is directly related to the size of the ejecta.  As the ejecta expand, the size of the emitting surface area increases and so does the flux density (e.g., Hjellming 1996; Seaquist \& Bode 2008).  In contrast, the radio light curves for embedded novae are dominated by synchrotron radiation in the first $\sim$ weeks, and therefore have high flux densities early and fade as the ejecta cool (e.g., O'Brien et al. 2006).

\begin{figure*}
\includegraphics[width=7.0in]{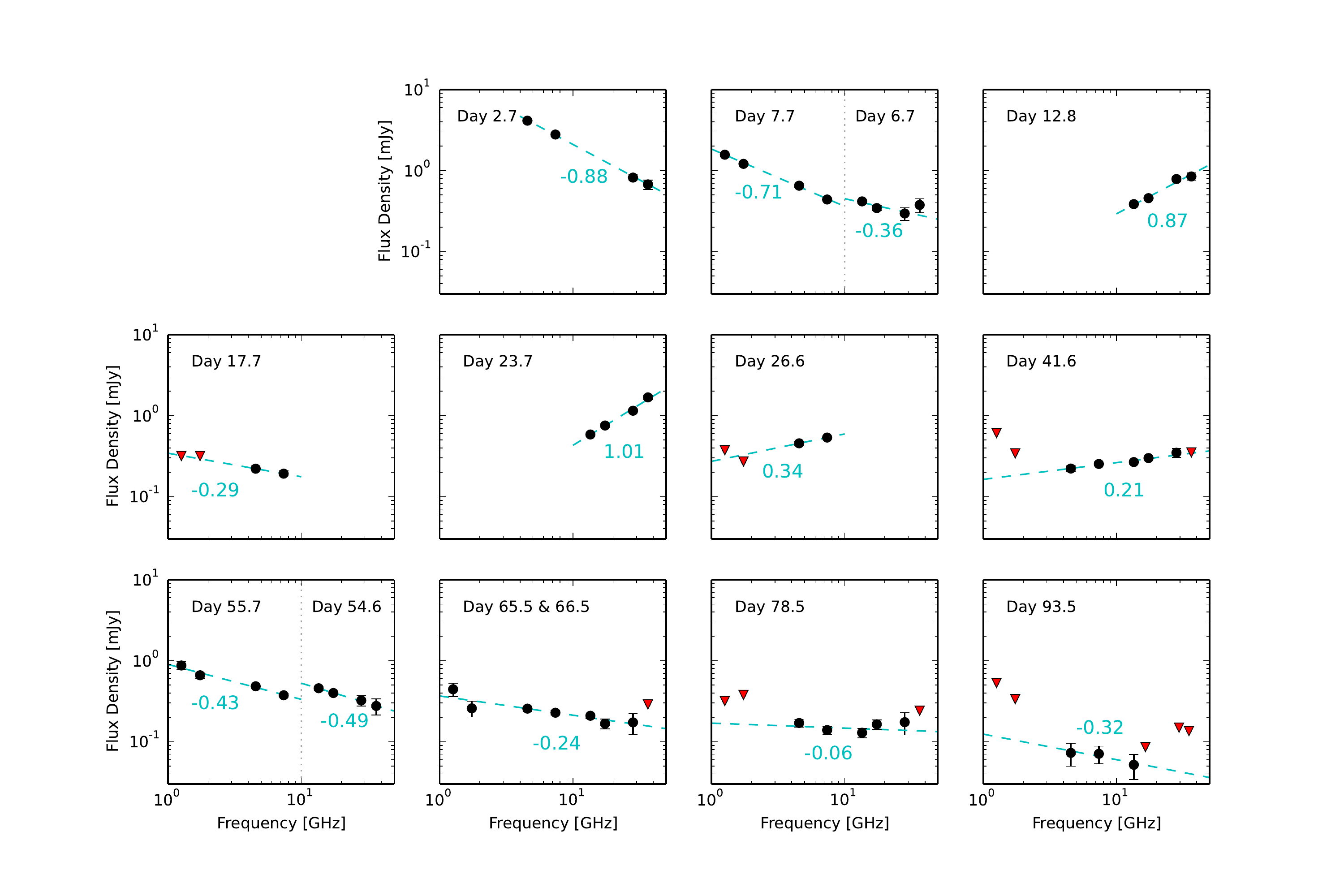}
\caption{Spectral energy distribution of VLA observations plotted for each epoch.  Non-detections are indicated by red downward-pointing triangles.  The fit to the same-day data is shown as a dashed cyan line, and the spectral index value is given in cyan.}
\label{vlasi}
\end{figure*}

We measured the spectral index $\alpha$ (using $S_{\nu}\propto\nu^{\alpha}$) for each of our VLA observations.  We fit the flux densities to a power law using the non-linear least squares \verb|curve_fit| function in the \emph{SciPy} package of python, weighted by the 1$\sigma$ uncertainties.  We ignored upper limit values for this fit, using only solid detections. If two observations were separated by less than 24 hours, they were combined to make a single spectral index measurement.  Our calculated spectral indices are presented in Table~\ref{vlasitab}, including the associated 1$\sigma$ uncertainties from the variance-covariance matrix.
The evolution of the spectral index is shown graphically in Figure~\ref{vlasi}.  During the first week of its evolution, the nova's radio spectral index rises toward lower frequencies, an indicator of optically thin synchrotron emission.  The spectral index then switches to rising toward higher frequencies.  This is typical of sources emitting via optically thick thermal bremsstrahlung.  To our knowledge, such a dramatic switch between a synchrotron-like $\alpha$ to a thermal-like $\alpha$ has never been seen before.  However, the spectral index of V1324 Sco showed evidence of being negative at early times followed by a positive slope later (Finzell et al. 2017), and Eyres et al. (2009) reported the presence of both positive and negative spectral indices present simultaneously for RS Oph.  At the tertiary peak around Day 56, the spectral index again rises toward lower frequencies and appears to be optically thin synchrotron.  This apparent switch from a thermal-like to synchrotron-like spectral index is unprecedented.  As the nova fades, its spectral index flattens as expected for an optically thin thermal bremsstrahlung source.

The evolution of V1535 Sco's radio spectral index is very different from the majority of classical novae.  Typically, a the radio emission begins with a positive (rising toward higher frequencies) spectral index while the ejecta is optically thick.  As the ejecta becomes optically thin, the radio spectral index flattens to value around -0.1.  This transition from strongly positive to flat begins with the highest frequencies first and progresses through the lower frequencies as the ejecta dims overall.  Even in novae where shocks are detected such as V959 Mon, the spectral index at early times is still usually positive (Chomiuk et al. 2014).  A negative spectral index, especially during the first few weeks of a nova's evolution, usually occurs in systems with evolved companions producing a strong wind for the nova ejecta to shock against, as in RS Oph (e.g., Eyres et al. 2009).

\begin{deluxetable}{cccc}
\tablewidth{0 pt}
\tabletypesize{\footnotesize}
\setlength{\tabcolsep}{0.025in}
\tablecaption{ \label{vlasitab}
VLA Spectral Indices}
\tablehead{UT Date & Day\tablenotemark{1} & Freq. Range & Spectral Index \\
&  & (GHz) & }
\startdata
2015-02-14 & 2.663 & 4.5\textendash 36.5 & -0.88$\pm$0.02 \\
2015-02-18 & 6.663 & 13.5\textendash 36.5 & -0.36$\pm$0.22 \\
2015-02-19 & 7.663 & 1.26\textendash 7.38 & -0.71$\pm$0.02 \\
2015-02-24 & 12.762 & 13.5\textendash 36.5 & 0.87$\pm$0.10 \\
2015-03-01 & 17.663 & 4.55\textendash 7.38 & -0.29$\pm$0.34\tablenotemark{\dag} \\
2015-03-07 & 23.663 & 13.5\textendash 36.5 & 1.01$\pm$0.07 \\
2015-03-10 & 26.563 & 4.55\textendash 7.38 & 0.34$\pm$0.14\tablenotemark{\dag} \\
2015-03-25 & 41.583 \& 41.633 & 4.55\textendash 28.2 & 0.21$\pm$0.04 \\
2015-04-07 & 54.563 & 13.5\textendash 36.5 & -0.49$\pm$0.02\\
2015-04-08 & 55.733 & 1.26\textendash 7.38 & -0.43$\pm$0.05 \\
2015-04-18 \& 19 & 65.513 \& 66.493 & 1.26\textendash 28.2 & -0.24$\pm$0.13 \\
2015-05-01 & 78.463 \& 78.503 & 4.55\textendash 28.2 & -0.06$\pm$0.13 \\
2015-05-16 & 93.463 & 4.55\textendash 13.5 & -0.32$\pm$0.15 \\
\enddata
\tablenotetext{1}{We take the time of initial detection 2015 February 11.837 UT \\(MJD 57064.837) to be Day 0.0}
\tablenotetext{\dag}{Only 2 data points available for fit}
\end{deluxetable}

\subsection{VLBA Compact Emission}

\begin{figure}
\includegraphics[width=3.3in]{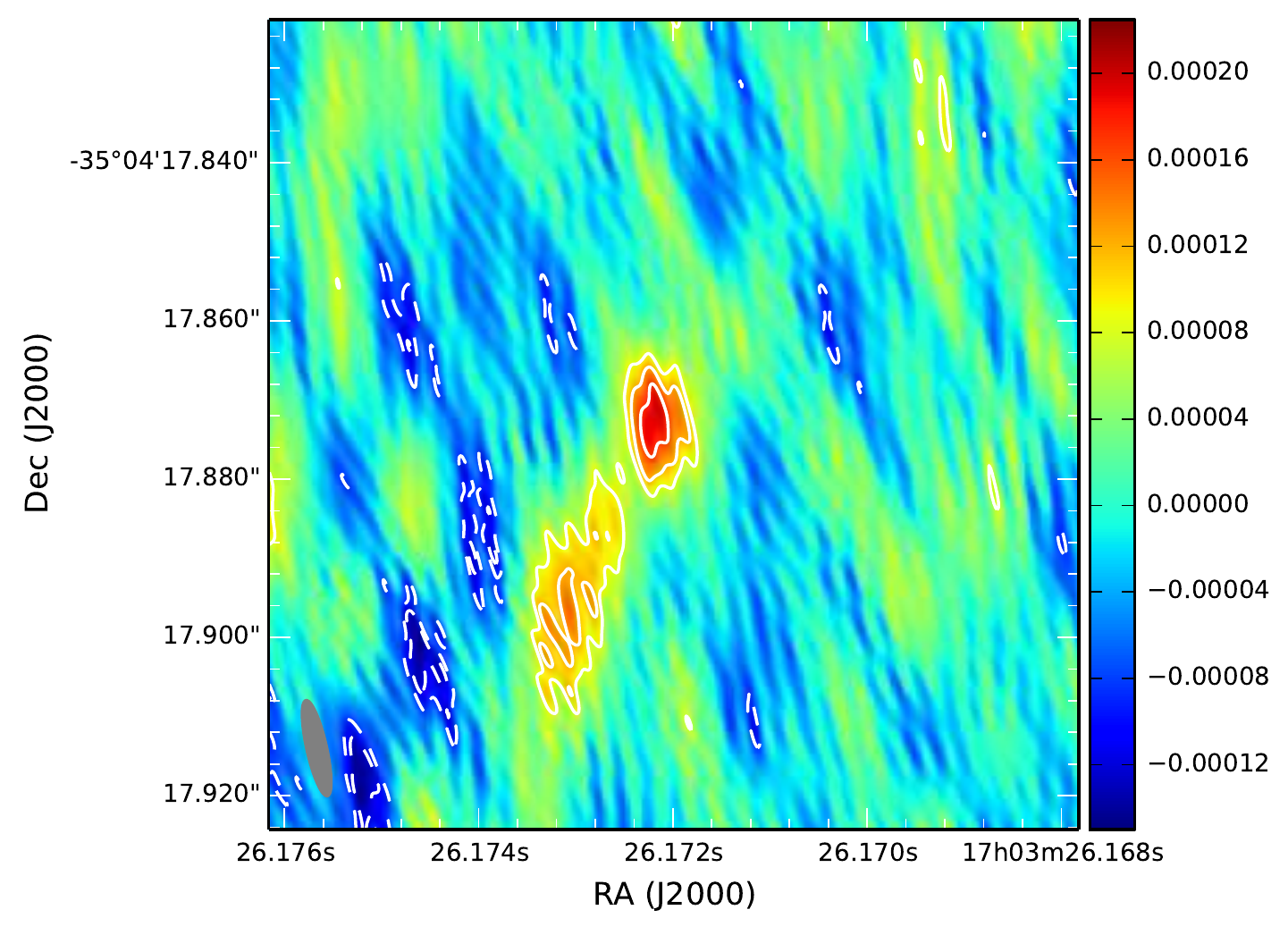}
\caption{4.87 GHz VLBA image of V1535 Sco on 2015 February 19 (Day 7.6).  The gray ellipse in the lower left corner indicates the shape and size of the restoring beam.  The apparent extension to the southeast is likely an imaging artifact resulting from imperfectly calibrated data.  The image is 0.1 arcsec on a side.}
\label{vlba1}
\end{figure}

Our VLBA detection on 2015 February 19 (Figure~\ref{vlba1}) was simultaneous with VLA L-band and C-band observations.  Using Difmap, we modelled the VLBA detection as a circular Gaussian in the \emph{uv}-plane.  The resulting total flux density is 0.477 mJy, with an off-source image rms of 0.044 mJy/beam, and a diameter of 6.52 mas.  The resolving beam for this observation was 2.5 x 12.4 mas, so this could indicate a resolved source.  

The brightness temperature (T$_{B}$) in K is given by the Rayleigh-Jeans relation (e.g., Rohlfs \& Wilson 2006)
\begin{equation}
S_{\nu} = \frac{2\, k_{B}\, T_{B}\, \Omega}{\lambda^{2}}
\end{equation}
where $S_{\nu}$ is the flux density in erg s$^{-1}$ cm$^{-2}$ Hz$^{-1}$ at frequency $\nu$, $\lambda$ is the observing wavelength in cm, $k_{B}$ is the Boltzmann constant in erg K$^{-1}$, and $\Omega$ is the solid angle of the source.  For a resolved source with a circular Gaussian shape, $\Omega = \pi\theta_{D}^{2}(4\ln2)^{-1}$, where $\theta_{D}$ is the angular diameter of the source in radians.  Therefore, the brightness temperature is calculated using
\begin{equation}
T_{B} = \frac{2\, \ln2\, S_{\nu}\, \lambda^{2}}{\pi\, k_{B}\, \theta_{D}^{2}}
\end{equation}
for our source.
This gives a brightness temperature of 5.8$\times$10$^{5}$ K.  This is significantly lower than the 10$^{7}$ K $T_{B}$ of the synchrotron components in RS Oph (Rupen et al. 2008).  However, we emphasize that our observation was made with lower resolution due to the drop-out of the Mauna Kea antenna.  It is possible that our single structure is actually composed of 2 or more components that are too compact to resolve with our configuration.  Modelling the source as two point sources results in a similar total flux density (0.445 mJy) and image rms (0.04 mJy/beam) with the same reduced $\chi^{2}$ (1.09) as the circular Gaussian model.  If we make the assumption that the angular size of the components is equal to the minor axis of the restoring beam (2.5 mas), the resulting $T_{B}$ estimates are 1.8$\times$10$^{6}$ K and 1.9$\times$10$^{6}$ K.  These are still lower than the $T_{B}$ for RS Oph, but are consistent with the $T_{B}$ for the compact components in V959 Mon (Chomiuk et al. 2014).

The simultaneous VLA flux density at 4.55 GHz was 0.650 mJy.  Because the VLBA is only sensitive to high brightness temperature (i.e., compact) emission, and the spectral index from the VLA observations of the nova ejecta was negative (i.e., larger flux density at lower frequencies), we conclude that the VLBA emission is most likely from synchrotron radiation.  The compact emission on this day dominated the total emission at the 73\% level.  The excess VLA radio flux on this day may be due to thermal bremsstrahlung emission from the shocked plasma or the pre-shocked, cooler outflow.

Using the VLBA flux density of the single circular Gaussian model component, we estimate the magnetic field strength using the revised equipartition formula from Beck \& Krause (2005).  This formula requires knowledge of the path length through which the radiation propagates $l$, the filling factor of the emitting material $f$, the inclination angle of the field $i$, and the proton-to-electron number density ratio, $K_{0}$: 
\begin{equation}
B\propto \left(\frac{K_{0}}{l\, f\, \cos(i)^{(1-\alpha)}}\right)^{1/(3-\alpha)}  
\end{equation}
where $\alpha$ is the radio spectral index ($S_{\nu}\propto\nu^{\alpha}$).
To make our estimate, we will assume a filling factor of 1, and an inclination angle of 0 (i.e., the field is face-on).  For $l$, we assume the emitting region was relatively thin,  $\sim10\%$ of the radius of the H$\alpha$ emitting region (see Figure~\ref{sco15_vnr}) at the time of the observation, which gives us $l\sim9\times10^{12}$ cm.  We used the typical values for $K_{0}$ of 40 and 100.  The resulting magnetic field strengths were 0.14 G for $K_{0}=40$ and 0.18 G for $K_{0}=100$.  If we assume that the path length is $\sim10\%$ of the radius of an ejecta shell expanding at a constant velocity of 1659 km s$^{-1}$ (the maximum velocity measured from the H$\alpha$ linewidth), we get $l\sim10^{13}$ cm and a magnetic field of 0.13 G for $K_{0}=40$ and 0.17 G for $K_{0}=100$.  If we assume that the path length is $\sim10\%$ of the radius of an eject expanding at a constant velocity of 4782 km s$^{-1}$ (the median velocity from the high velocity outliers in the H$\alpha$ spectra over the first 5.5 days), we get $l\sim3\times10^{12}$ cm and a magnetic field of 0.10 G for $K_{0}=40$ and 0.13 G for $K_{0}=100$.  These values are all larger than, but still comparable to, the magnetic field strengths estimated for RS Oph of $0.03(1+K_{0})^{2/7}$ (Rupen et al. 2008).  Assuming the same values of $K_{0}$ as we used, the RS Oph magnetic field strength was between 0.087 G and 0.11 G.  It should also be noted that Rupen et al. (2008) estimated the magnetic field for RS Oph over 20 days after the eruption, whereas we are estimating the field for V1535 Sco only 7.7 days after eruption.  It is possible that the magnetic field declines over time as the ejecta expand and the shocks dissipate.

\subsection{Optical and Near-Infrared Photometry}
\begin{figure}
\includegraphics[trim=30 20 0 20,clip,width=3.5in]{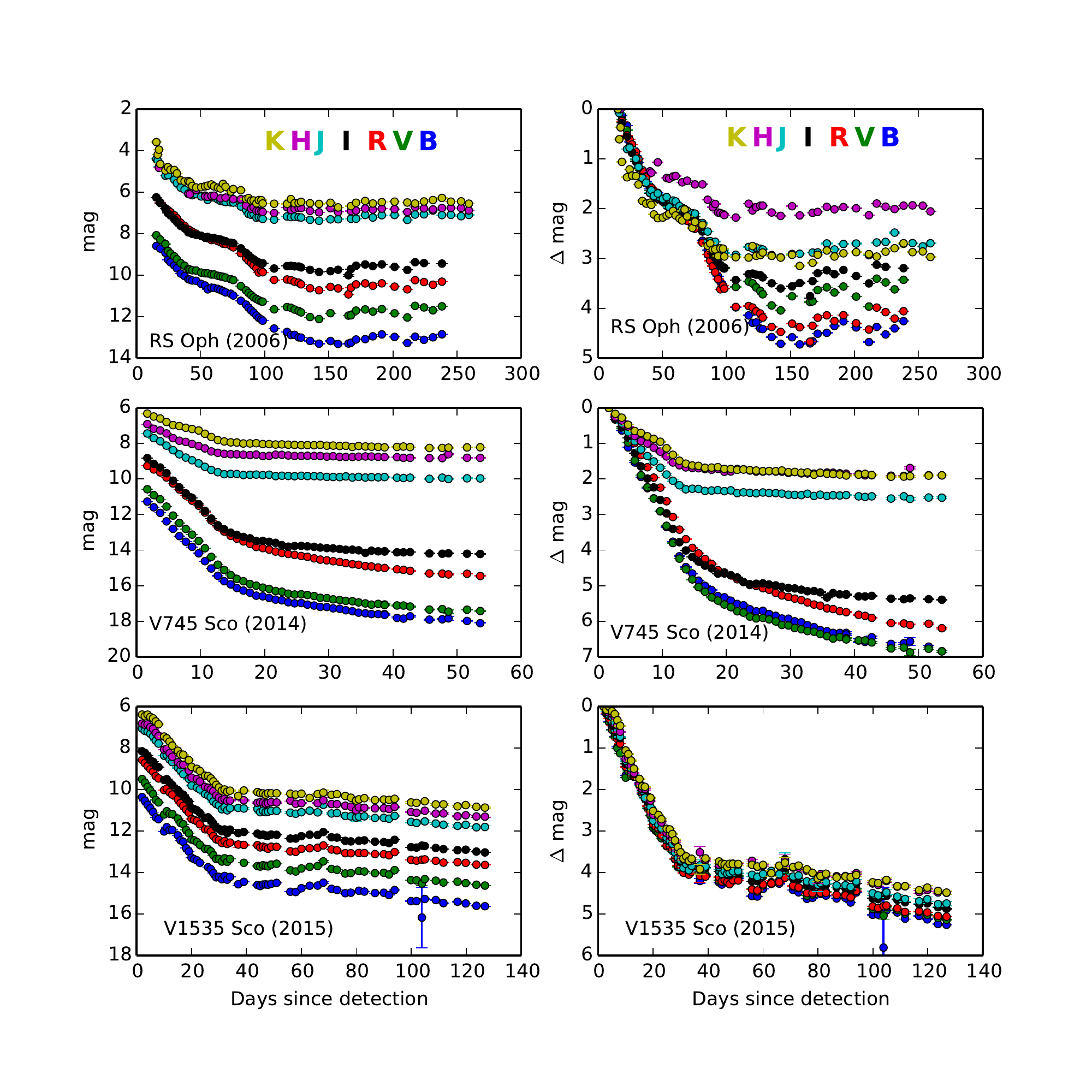}
\caption{Comparison of photometric measurements for two novae known to have red giant companions with V1535 Sco.  The left hand column shows the optical and near infrared light curves.  The right hand column shows the difference in magnitude from the first observation.  The first row is RS Oph (2006 eruption), the second row is V745 Sco (2014 eruption), and the bottom row is V1535 Sco.}
\label{smarts_comp} 
\end{figure}

The optical and near-infrared light curve for V1535 Sco is not consistent with light curves for other novae known to have red giant companions (see Figure~\ref{smarts_comp}).  In particular, when one considers the difference between the peak magnitude and the current magnitude (right hand side of Figure~\ref{smarts_comp}), V1535 Sco is clearly an outlier.  In other novae with red giant companions, the companion star begins to dominate the difference light curve at longer wavelengths relatively early, while the shorter wavelengths are still dominated by the nova ejecta.  V1535 Sco, on the other hand, does not follow this behavior.  Instead, all wavelengths fade together for the duration of the light curve.  This indicates that the companion star is unlikely to be a red giant.  However, the presence of synchrotron emission in the radio (Sections 4.1 and 4.2), the hard X-ray emission, and the rapid fading of the optical light curve all indicate the nova ejecta were shocking against an external medium.  

\begin{figure}
\includegraphics[width=3.25in]{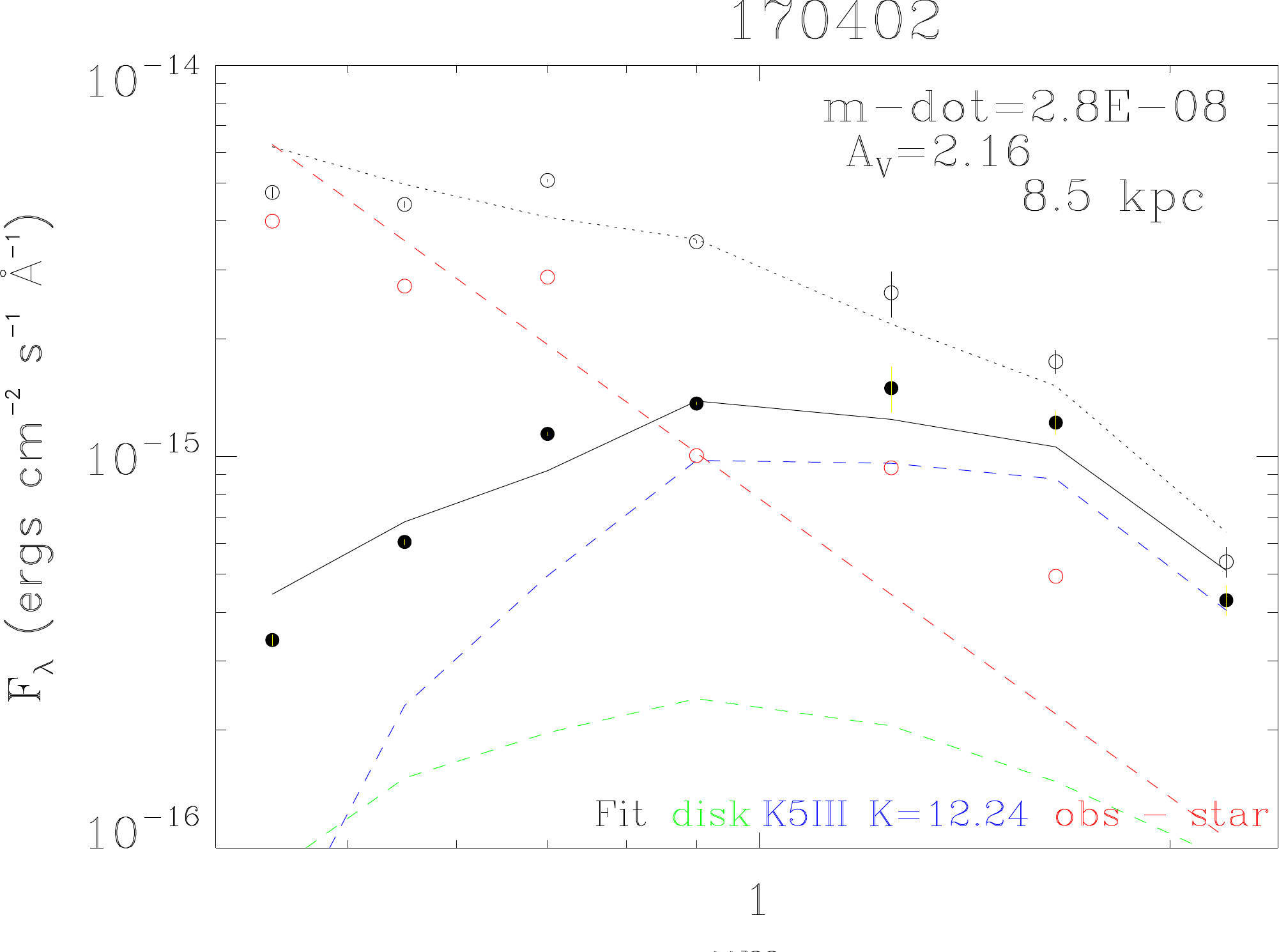}
\caption{The fit to the SED on 2017 April 2 with an active accretion disk and a K5~III
star. Model details are described in the text. Filled circles are the observed
data. The open white circles are the dereddened observations; open red circles
are the data after subtracting the K5 III SED. The green and blue dashed
curves are the fits to the accretion disk and the star, respectively. The red
dashed curve is the dereddened fit to the accretion disk. The fit underpredicts
the observed $J$ and $H$ fluxes.}
\label{smarts_sedfit} 
\end{figure}

One possible explanation is that the companion star is not an M giant, but a K giant.  We tested this hypothesis by fitting our optical and near infrared fluxes with a model that combined a hot accretion disk and a stellar spectral energy distribution for a K giant star. 

The quiescent unreddened optical colors of V1535 Sco are close to 0. We modelled the optical-infrared (OIR) spectral energy distribution (SED) as the sum of a mid-K giant plus an active accretion disk. The accretion disk
is constructed following the Bertout, Basri, \& Bouvier (1988) formalism.
The optically thick disk is constructed of a series of annuli, each of which
emits as a black body at a temperature set by its distance from the star. 
We set the inner edge of the accretion disk to the radius of a white dwarf;
exact details are not important since the boundary layer and the inner edge
of the disk emit in the ultraviolet. On the red side of the peak, the accretion disk
follows a power law F$_{\lambda} \sim \lambda^{-2.3}$.
With 7 OIR fluxes, the model is under-constrained. We fix the extinction A$_V$ to
be 2.16~mag and set the distance to 8.5~kpc. We constrain the white dwarf to be
hot. We constrain the donor star to be K5~III, with VJHK colors taken from
Koornneef (1983). Free parameters in the fit are the magnitude of the donor
star and the mass accretion rate.  An example of a fit is shown in Figure~\ref{smarts_sedfit}.

We fit the ten SEDs obtained 2016 April 25 through 2017 April 02. Over that 
year the donor star $K$ faded by 0.15$\pm$0.07 mag. This fading could be a
cooling after irradiation by the nova, or it could be part of a longer term
variation. The mean fit $K$ of 12.14 is consistent with the 2MASS $K$ magnitude.
Meanwhile the mass accretion rate has dropped linearly by a factor of 6, from
12 to 2$\times$10$^{-8}$M$_\odot$~yr$^{-1}$. Numbers should be taken with a
grain of salt. The mass accretion rate and the assumed distance are strongly
correlated where the inferred mass accretion rate scales as the cube of the inferred distance. 

Using our fits and assuming a distance of 8.5 kpc, we find that the absolute $J$ magnitude for V1535 Sco in quiescence should be -1.2.  Covey et al. (2007) gives the absolute $J$ magnitude for a K3~III star as -1.13, which is remarkably close for such a simple model.  Munari et al. (2017) also conclude that the companion in V1535 Sco is consistent with a K3-4 III giant.

\subsection{Optical and Near Infrared Spectroscopy}
\begin{figure}
\includegraphics[width=3.5in]{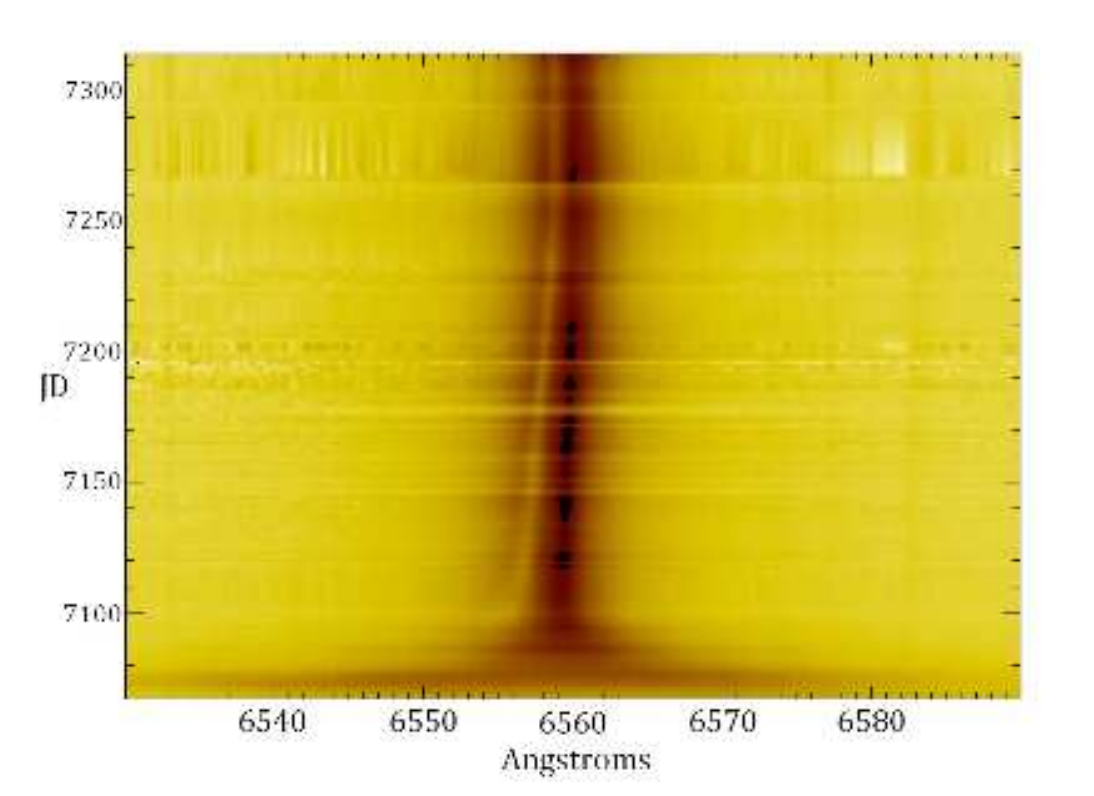}
\caption{Negative image of the H$\alpha$ region of the spectrum as a function of time.  Darker regions indicate more emission, brighter regions indicate absorption.}
\label{walterha}
\end{figure}
Srivastava et al. (2015) presented measurements of the ejecta velocity for V1535 Sco based on the Pa$\beta$ emission line at 1.2818 $\mu$m starting 7 days after the detection and ending 40 days after the detection.  Their velocities are given as full width at half maximum (FWHM) of the line.  As with our H$\alpha$ measurements, we approximate the FW3$\sigma$ velocities by applying the same calculation as described in Section 2.4.  We then fit the line-width-vs-time data with a power law using the non-linear least squares \verb|curve_fit| technique in the \emph{SciPy} package of python.    
Our resulting fit is somewhat different from that given in Srivastava et al. (2015).  They report a power law of $t^{-1.13\pm0.17}$ whereas our fit gives $t^{-0.89\pm0.14}$.  However, we should note that we are uncertain as to the methods used by Srivastava et al. (2015) because they do not describe their fitting procedure in detail.  To compare the Pa$\beta$ results with our more densely sampled H$\alpha$ data, we fit our H$\alpha$ FW3$\sigma$ measurements from Day 7.48 to Day 39.4 using the same \emph{SciPy} method and find a power law of $t^{-1.49\pm0.05}$, which nearly agrees within uncertainties with the Srivastava et al. (2015) value.  Of course, this simple power-law fit completely ignores the complicated behavior observed during the first week following the explosion (see Figure~\ref{sco15_vnr}).

The SMARTS optical spectroscopic monitoring of V1535 Sco also revealed blue-shifted H$\alpha$ absorption features.  These features become noticeable around Day 15, after the H$\alpha$ line has narrowed significantly (see Figure~\ref{walterha}).  The blue-shifted features have an initial velocity of $\sim500$ km s$^{-1}$, but slow to $\sim50$ km s$^{-1}$ by Day 247.  These absorption features indicate the presence of a large amount of cool, neutral material ahead of the H$\alpha$ emitting region.  It is possible that this neutral material was pre-shocked wind material being swept up by the nova ejecta in a ``snowplow'' fashion.

\subsection{X-rays}
We fit the \emph{Swift} data using the \verb|XSpec| version 12.8.2 package.  The spectra were best fit using a combination of a black body (BB) and a thermal plasma (apec).  The apec model includes contributions from free-free continuum and lines (Smith et al. 2001).  Only the first 5 detections had high enough signal-to-noise to successfully model the emission.  The results are given in Table~\ref{swiftab2}.  The Hydrogen column density (N$_{\rm H}$) appears to be fairly constant throughout the first 18 days, with a deviation on Day 11 which is most likely the fitting software getting stuck in a local minimum.  This constant N$_{\rm H}$ is quite different from the sharply decreasing N$_{\rm H}$ observed in the embedded novae RS Oph (Bode et al. 2006) and V745 Sco (Page et al. 2015).  The black body temperature appears to decrease by only a factor of 2 over the first 25 days, with a minor fluctuation around Day 17.9.  The hot plasma temperature, on the other hand, dramatically cools by at least 2 orders of magnitude during the same time period.  This is likely due to the shocks expanding and becoming radiatively efficient.

The $E(B-V)$ of 0.96 reported by Munari et al. (2017) implies N$_{\rm H}\approx5\times10^{21}$ cm$^{-2}$ (Predehl \& Schmitt 1995).  Our values of N$_{\rm H}$ are nearly twice this, indicating a local enhancement from a possible stellar wind.  Interestingly, our N$_{\rm H}$ values are quite similar to those measured for 2 recurrent novae known to have red giant companions: V745 Sco and RS Oph.  In RS Oph, the values for N$_{\rm H}$ were $\sim10^{23}$ cm$^{-2}$ just 0.16 days after eruption, but fell to $\sim10^{22}$ cm$^{-2}$ by Day 3.23 (Page et al. 2015) and were $\sim 8\times10^{21}$ cm$^{-2}$ by Day 10 (Orio et al. 2015).  In RS Oph, Nelson et al. (2008) reported N$_{\rm H}\sim10^{22}$ cm$^{-2}$ 14 days after eruption.  We stress that this does not directly indicate the presence or absence of a red giant, especially as the kT values for V1535 Sco are similar to those that Mukai \& Ishida (2001) found for V382 Vel, which is known to have a main sequence companion.  However, it is important to note that V1535 Sco was detected by \emph{Swift} only 5 days after its discovery in the optical, while V382 Vel was not detected in X-rays until 20 days after optical discovery.  The delay in X-rays in V382 Vel is often seen in non-embedded novae, and is thought to be due to the fact that it takes time for multiple outflows to be ejected and then collide.  It may therefore be more appropriate to compare the X-ray emission from V382 Vel to the later (\textgreater Day 49) X-rays from V1535 Sco.

Overall, the X-ray emissions are somewhat puzzling.  On the one hand, the flat N$_{\rm H}$ for several days is different from the behavior in known embedded novae.  On the other hand, there appears to be local absorption in the system on levels similar to known embedded novae.  There are two other major points to consider.  First, the X-ray emission in V1535 Sco was hard early in its evolution, which is much more typical of embedded novae.  Second, those early X-rays were very bright.  Using our assumed distance of 8.5 kpc, the early kT of 40 keV, and N$_{\rm H}\approx 10^{22}$ cm$^{-2}$, we find an X-ray luminosity of $\sim4\times10^{35}$ erg s$^{-1}$, which is much higher than any non-embedded nova we are aware of.  For comparison, the X-ray luminosity for V382 Vel was $7\times10^{34}$ erg s$^{-1}$ (Mukai \& Ishida 2001), while RS Oph as over $10^{36}$ erg s$^{-1}$ (Mukai et al. 2008).  While the X-ray data do not provide proof that the nova was embedded in wind material from the companion star, it is at least a plausible explanation for the interesting behavior.


\section{Discussion}

\begin{figure*}
\includegraphics[trim=20 80 20 80,clip,width=7.5in]{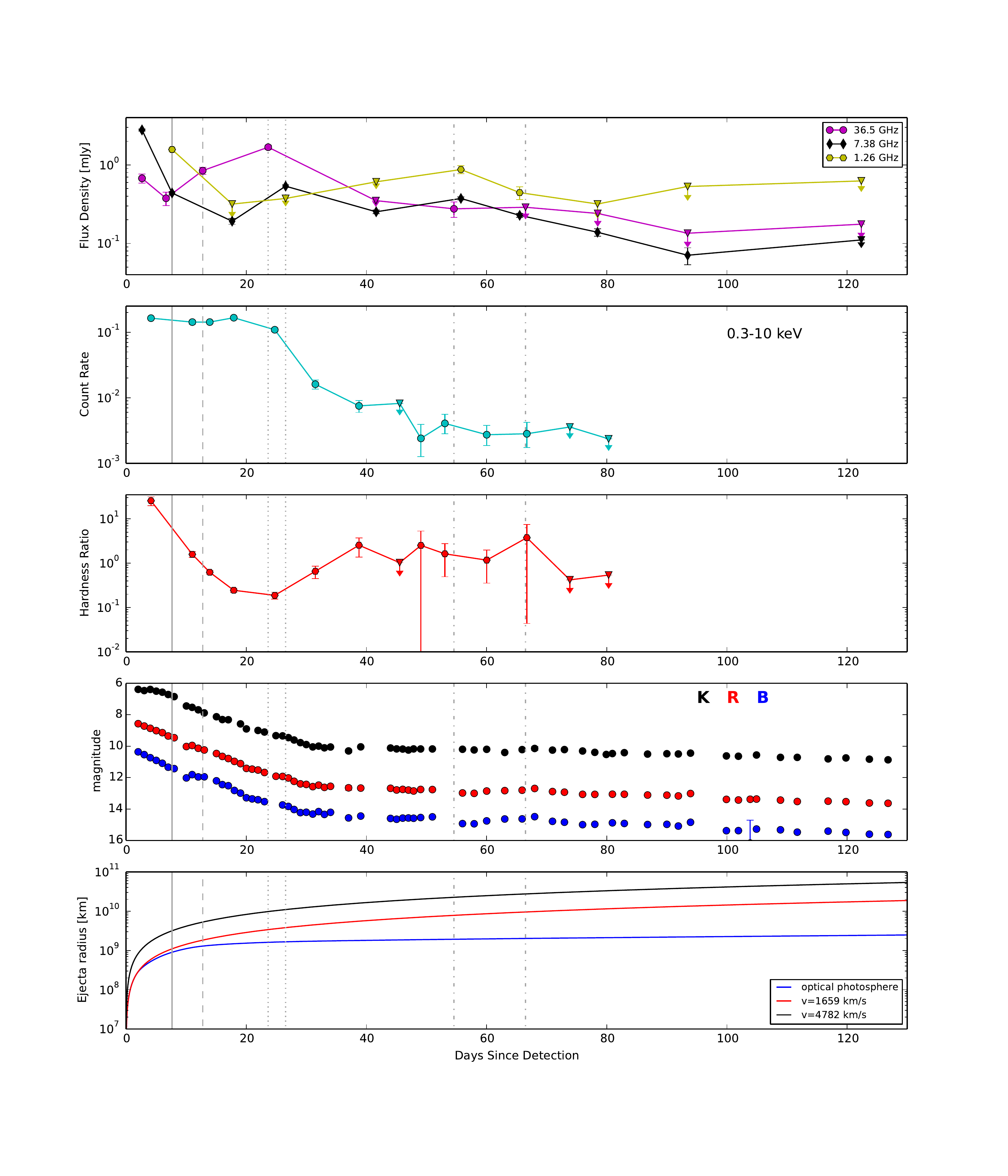}
\caption{{\bf Top:} VLA light curve at 36.5, 7.38, and 1.26 GHz. {\bf 2nd:} \emph{Swift} total count rate. {\bf 3rd:} \emph{Swift} hardness ratio (1-10 keV/0.3-1.0 keV).  {\bf 4th:} Optical photometry from the SMARTS program. {\bf Bottom:} Estimate of the radius of the ejecta {\bf shell}: blue -- velocity from H$\alpha$ line widths; red -- constant velocity of 1659 km s$^{-1}$; black -- constant velocity of 4782 km s$^{-1}$.  {\bf Vertical Lines:} Solid = VLBA detection; Dashed = VLBA non-detection; Dotted = secondary maximum in the radio light curve; Dash-dotted = period of apparent second synchrotron bump in VLA spectral indices.}
\label{rxofig}
\end{figure*}

Bringing together all of our various observations and examining them together leads to some interesting conclusions.  Figure~\ref{rxofig} shows information from all of our observing wavelengths with important times (e.g., the VLBA detection and the second radio synchrotron emission event) indicated.  The initial negative radio spectral index combined with hard X-rays indicates synchrotron radiation produced by strong shocks.  This is expected for a nova exploding into a thick wind from a giant companion.  The transition from a negative to positive radio spectral index corresponds with both a softening of the X-rays while the total X-ray count rate remains relatively constant.  This indicates the shocks are weakening and the plasma is cooling.  At the same time, the radio thermal photosphere is expanding and begins to dominate the radio emission.  
\begin{center}
\begin{deluxetable*}{cccccccc}
\tablewidth{0 pt}
\tabletypesize{\footnotesize}
\setlength{\tabcolsep}{0.025in}
\tablecaption{ \label{swiftab2}
Best-Fit Model Parameters from \emph{Swift} Observations}
\tablehead{Obs\tablenotemark{1} & Day\tablenotemark{2} & N$_{\rm H}$ & kT$_{BB}$ & norm$_{BB}$ & kT$_{apec}$ & norm$_{apec}$ & cstat/DoF \\
ID &  & ($10^{21}$ cm$^{-2}$) & (eV) &  & (keV) &  & }
\startdata
2 & 4.16 & 9.5$\pm$1.9 & 67$\pm$11 & 349$^{+14210}_{-341}$ & \textgreater41 & 0.021$\pm$0.003 & 373.39/377 \\
3 & 11.04 & 3$\pm$2 & 45$^{+16}_{-11}$ & 0.0037$^{+0.1359}_{-0.0035}$ & 3.8$^{+2.7}_{-1.3}$ & 0.0027$^{+0.0007}_{-0.0005}$ & 214.39/121 \\
4 & 13.94 & 9.6$\pm$1.9 & 35$\pm$4 & 5.9$^{+57.7}_{-5.3}$ & 1.0$\pm$0.09 & 0.0037$^{+0.0011}_{-0.0009}$ & 151.98/159 \\
5 & 17.93 & 9.61$\pm$2.4 & 45$^{+8}_{-4}$ & 0.74$^{+7.5}_{-0.68}$ & 1.013$^{+0.22}_{-0.17}$ & 0.0018$^{+0.0006}_{-0.0004}$ & 203.73/138 \\
6 & 24.75 & 18.4$^{+5.9}_{-8.3}$ & 31$^{+17}_{-45}$ & 4264$^{+21089}_{-4240}$ & 0.27$^{+0.65}_{-0.09}$ & 0.026$^{+0.200}_{-0.025}$ & 80.10/88 \\ 
\enddata
\tablenotetext{1}{All Observation ID's are 0003363400X}
\tablenotetext{2}{We take the time of initial detection 2015 February 11.837 UT (MJD 57064.837) to be Day 0.0}
\end{deluxetable*}
\end{center}

The ejecta cool and fade for several weeks until around Day 49 when the X-ray count rate appears to increase slightly (going from a non-detection to detection again) and the radio spectral index again shows evidence of synchrotron emission.  There are two possible explanations for this second episode of synchrotron emission.  First, the ejecta encountered another dense medium to shock against.  Using the fit to the velocity from H$\alpha$ emission lines and integrating over time, the H$\alpha$ emitting region had a radius of approximately 13 AU by this time.  If we assume that the ejecta that were emitting synchrotron radiation at this time was instead travelling at a velocity of 1659 km s$^{-1}$ for 49 days, that would be at a radius of approximately 47 AU.  If we assume that the synchrotron-emitting material was from the ejecta producing the high velocity outliers in the H$\alpha$ spectra and travelled with a constant velocity of 4782 km $^{-1}$, it would have travelled 135 AU by Day 49.  It is possible that the ejecta caught up to material ejected during a previous nova eruption.  If we assume that the slowest material from a previous ejecta were moving at 50 km $^{-1}$, the previous eruption would have occurred 1.2 years ago for a distance of 13 AU, 4.5 years ago for a distance of 47 AU, or 12.8 years ago for a distance of 137 AU.  Of these three, the 12.8 years seems the most likely, as it is hard to imagine the astronomy community could have missed multiple outbursts of this nova in the past 10 years.  However, a recurrence time of 12.8 years would still require that this system is a very young recurrent nova undergoing only its second or third outburst, or that we have missed a very large number of its previous eruptions.

The second possibility is that the second synchrotron event is the result of shocks within the ejecta itself.  We already have evidence for two different outflows from the width of the H$\alpha$ line and the high velocity outliers (see Section 2.4).  It is not hard to imagine an interaction between these two outflows, or possibly even a third outflow such as a hot, fast wind driven by thermonuclear burning on the surface of the white dwarf but hidden below the optical photosphere of the bulk of the ejecta.  

\subsection{Expected Brightness Temperature for VLBA Source}
Using the radial size, we can estimate what we expect the brightness temperature of the ejecta to be at the time of the VLBA observation using
\begin{equation}
T_{B} \approx \frac{S_{\nu}\, c^{2}\, D^{2}}{2\, \pi\, k_{B}\, \nu^{2}\, R^{2}}
\end{equation}
where $R$ is the radius of the spherically expanding ejecta and $D$ is the distance to the source (e.g., Seaquist \& Bode 2008).  We will start by assuming the size of the radio emitting region is the same as the H$\alpha$ emitting region.  From our fit to the H$\alpha$ line widths presented in Section 2.4, we derive a radius for the H$\alpha$ emitting region of $8.98\times10^{13}$ cm on Day 7.663.  Using only the VLBA flux of 0.477 mJy, and assuming a distance of 8.5 kpc, this gives $T_{B} \approx 1.8 \times 10^{7}$ K.  Alternatively, the radius could be the distance travelled by ejecta moving with a velocity of 1659 km s$^{-1}$, which gives is $1.10\times10^{14}$ cm.  Using the VLBA flux and a distance of 8.5 kpc, this material would have a $T_{B} \approx 1.2 \times 10^{7}$ K.  Both of these are significantly higher than the estimate of $5.8\times10^{5}$ K using the distance-free formula for $T_{B}$ in Section 4.2.  To get the largest possible radius, we could assume the ejecta were travelling at 4782 km s$^{-1}$ for 7.663 days, giving a radius of $3.17\times10^{14}$ cm and $T_{B} \approx 1.4 \times 10^{6}$ K.  The discrepancy between the observational and expected $T_{B}$ can indicate several possibilities: first, the ejecta were unresolved by the VLBA; second, the nova is much closer than 8.5 kpc; third, the ejecta shape is highly non-spherical; or some combination of these reasons.  We should note that the expected $T_{B}$ using the largest possible radius agrees well with the observational $T_{B}$ assuming 2 point sources.  However, using the largest possible radius would only give us a single sphere with an angular size of 2.5 mas at 8.5 kpc (which just happens to be the minor axis of the VLBA restoring beam for our observation), not the 2 point sources modelled in Section 4.2, so we cannot reconcile the measurement with the observation quite so easily.

From Section 3, recall that the closest the nova can be is 5.8 kpc.  At this distance, the brightness temperature would be $8.3 \times 10^{6}$ K for a radius of $8.98\times10^{13}$ cm, $5.5 \times 10^{6}$ K for a radius of $1.10\times10^{14}$ cm, or $6.7 \times 10^{5}$ K for a radius of $3.17\times10^{14}$ cm.  The first two are still larger than our value of $5.8 \times 10^{5}$ K we calculated for the single-component model in Section 4.2, although they are comparable to the two-component model.  The final one, using our maximum possible radius, agrees well with the single-component model $T_{B}$ in Section 4.2.  However, recall that we find it more likely that the high velocity outliers in the H$\alpha$ spectra represent a bipolar outflow, not an expanding spherical shell.

Note that this approach to estimating what the brightness temperature should be at the time of the VLBA detection assumes that the entire surface of the ejecta shell is emitting.  If the emission is coming from compact, jet-like structures as in RS Oph (Rupen et al. 2008; Sokoloski et al. 2008), the value for $R$ would be smaller and the estimate for $T_{B}$ would be even larger.  Placing the nova at further distances also increases the estimate for $T_{B}$.

Because placing the nova closer does not make the values of $T_{B}$ match, it is more likely that the VLBA did not fully resolve the compact emitting region on Day 7.663.  Because of the highly-elliptical shape of the restoring beam, it is especially likely that there is unresolved structure in the north-south direction.  However, the image does look like there is some resolved structure in the east-west direction.  We find it likely that the compact component is non-spherical, possibly bi-lobed, with more extension in the east-west direction.

\subsection{X-ray and Radio Emission Measures Near Day 25}
We used the emission measures for both radio and X-ray to investigate whether the secondary maximum in the radio light curve that occurs between Day 23.663 and 26.563 can be explained by thermal radio emission from the X-ray emitting plasma.  For the X-rays, the emission measure depends on the volume of the emitting region:
\begin{equation}
EM_{x} = \int n_{i}\, n_{e}\, dV
\end{equation}
where $n_{i}$ and $n_{e}$ are the number densities of ions and electrons, respectively, and $V$ is the volume of emitting material.
Generally it is assumed that $n_{e} \approx n_{i} = constant$ so that it simplifies to $EM_{x} \approx n_{e}^{2}V$.  To make a further simplification, the emitting volume is assumed to be thin compared to the total extent of the ejecta so that it can be approximated as a uniform slab: $V \approx \pi r^{2} dl$, where $r$ is the radius of the ejecta (we are assuming the emitting region can be approximated as a cylinder) and $dl$ is the thickness of the emitting region.  The X-ray emission measure can be determined from the hot plasma parameters in the fits to the \textit{Swift} data: 
\begin{equation}
EM_{x} = (4 \times 10^{14})\, \pi\, D^{2}\, norm_{apec}
\end{equation}
where $D$ is the distance to the nova.  Assuming a distance of 8.5 kpc (placing the nova approximately at the Galactic Center) and using $norm_{apec}$ from Day 24.75 (see Table~\ref{swiftab2}) gives $EM_{x} \approx 2.25 \times 10^{58}$ cm$^{-3}$.  Assuming the ejecta travel with a constant velocity of 1659 km s$^{-1}$, we can determine the approximate radius for the shell on Day 24.75 was $3.5 \times 10^{14}$ cm.  If we assume that the thickness of the emitting shell is $\sim10\%$ of its radius, we can solve for the number density of the emitting electrons: $n_{e} \approx 4.0 \times 10^{7}$ cm$^{-3}$ emitting at a temperature of approximately $10^{6}$ K (from kT$_{apec}$ in Table~\ref{swiftab2}).  Note that if we assume a larger distance to the nova, both $EM_{x}$ and $n_{e}$ increase.

The radio (path length) emission measure is given by:
\begin{equation}
EM_{r} = n_{e}^{2}\, dl
\end{equation}
plugging in the same $dl$ (but this time in pc) used for $EM_{x}$ and using $n_{e}$ derived from the X-ray data gives $EM_{r} = 1.8 \times 10^{10}$ cm$^{-6}$ pc.  The radio emission measure is also part of the equation for optical depth, $\tau_{\nu}$ (Rohlfs \& Wilson 2006):
\begin{equation}
\tau_{\nu} = 8.235\times10^{-2}\, T_{e}^{-1.35}\, \nu^{-2.1}\, EM_{r}\, a(\nu,T)
\end{equation}
Therefore, another way to calculate the radio emission measure is given by:
\begin{equation}
EM_{r} = 12.143\, T_{e}^{1.35}\, \nu^{2.1}\, \tau_{\nu}\, a(\nu,T)^{-1}
\end{equation}
where $T_{e}$ is the electron temperature in K, $\nu$ is the observing frequency in GHz, $\tau_{\nu}$ is the optical depth at the observing frequency, and $a(\nu,T)$ is a correction term usually assumed to be 1.  In order for the emission to be optically thick at the observed frequency, $\tau_{\nu}\gtrsim1$.  Because we are trying to determine if the radio flux can be explained by the X-ray plasma, we set $T_{e} \approx 10^{6}$ K from the kT$_{apec}$ value from the fits to the Swift data on Day 24.75 (see Table~\ref{swiftab2}).  We use our highest observing frequency of 36.5 GHz because it provides the tightest constraint.  This gives us $EM_{r} \approx 2.92 \times 10^{12}$ cm$^{-6}$ pc, which is more than two orders of magnitude larger than expected from the X-ray electron number density used above.  Therefore, under the reasonable assumptions we have made, the X-ray emitting plasma cannot account for the thermal radio emission detected around Day 24.  It is more likely that there is also a warm ($\sim10^{4}$K) ionized ejecta that begins to dominate the radio emission at this time.

Unfortunately, the uncertainty on $norm_{apec}$ from Day 24.75 is very high.  If we use the $norm_{apec}$ value from Day 17.93 with its much lower uncertainty and assume that the number density remains mostly constant until Day 24, we get $n_{e} = 1.7\times10^{7}$ cm$^{-3}$, which would imply $EM_{r} = 2.4\times10^{9}$ cm$^{-6}$ pc.  This makes the discrepancy between the two methods for determining the radio emission measure even greater, further strengthening our argument that the X-ray emitting plasma cannot account for the radio flux density observed during the second radio maximum.

\section{Conclusions}

V1535 Sco showed peculiar behavior at nearly every wavelength we observed.  The radio emission started bright, faded quickly, and then had two re-brightening events.  The spectral index for early radio emission and the second radio re-brightening event were consistent with optically thin synchrotron emission.  The X-ray emission began promptly with a hard spectrum.  There was a re-brightening in the X-rays which appeared to correspond to the second re-brightening event in the radio.  The lack of connection between the first radio re-brightening event and the X-ray emission implies that are are at least two emitting components in the ejecta: one shock-heated plasma and one thermal bremsstrahlung.  The optical observations indicate that the nova was discovered post-peak, and faded very fast.  Optical spectroscopy also indicated the presence of two outflows: 1) a relatively slow ($\sim$1659 km s$^{-1}$) outflow; and 2) a fast ($\sim$4782 km s$^{-1}$), possibly bipolar outflow.  Spectral monitoring of the H$\alpha$ line at late times also indicated the presence of a neutral, dense surrounding cloud of emitting material.


The early hard X-rays combined with the radio synchrotron emission and the detection with the VLBA strongly support the existence of strong shocks very early in the evolution of the nova.  Such early strong shocks are most easily explained by the presence of a dense wind from the companion star.  The estimated magnetic field from the VLBA detection was between 0.10 G and 0.18 G, which is larger than but comparable to the magnetic field strength found by Rupen et al. (2008) for RS Oph, a system known to have a red giant companion.

There was evidence for a second shock around Day 50.  The radio spectral index at this time changed from being consistent with optically thick thermal bremsstrahlung, to being synchrotron-like.  The X-ray emission also showed an increase at this time.  We posit that this second shock was the result of collisions between multiple outflows within the ejecta, but the presence of a shell of dense material (possibly from a previous eruption) has not been completely ruled out.


The nova had strong hard X-ray emission early in its evolution, but no detectable $\gamma$-rays.  However, it should be noted that V1324 Sco is the only nova \emph{Fermi} has detected which has distance comparable to V1535 Sco (Finzell et al. 2015).  We also note that V745 Sco, with its red giant companion and presumably denser wind, was only marginally detected by \emph{Fermi} (Cheung et al. 2014).  It is very likely that V1535 Sco produced at least some $\gamma$-ray emission, but it was simply too far away for \emph{Fermi} to detect.

To date, only a handful of Galactic novae with red giant companions are known, including the recurrent novae RS Oph and V745 Sco.  However, recent studies on the nova population in M31 indicate that there may be many novae with red giants that are not detected, and they may constitute $\sim$30\% of the nova eruptions in the Milky Way (Williams et al. 2016).  There is particular interest in discovering more of these novae because they are possible progenitors to Type Ia supernovae (e.g., Starrfield et al. 2012).  In fact, Dilday et al. (2012) claim that the supernova PTF 11kx originated from such a system.  

Novae that occur in a system with a red giant companion are often referred to as ``symbiotic novae''.  However, the term  ``symbiotic novae'' is also used to describe a particularly slow and long-lasting class of novae where nuclear burning on the surface of the white dwarf is sustained for several years, and sometimes decades such as PU Vul, HM Sge, and AG Peg (e.g., Iben Jr. \& Fujimoto 2008).  In order to avoid confusion and better describe the general class of thermonuclear novae with evolved companions, our collaboration has adopted the term ``embedded nova'' to refer to any nova that is embedded in the dense wind of its post-main sequence companion star, regardless of the duration of its optical maximum (e.g. Chomiuk et al. 2012; Mukai et al. 2014).

We find some evidence that V1535 Sco is an embedded nova.  The early strong, hard X-rays and non-thermal radio emission argue for the presence of a pre-existing dense material to shock against.  Also, the X-ray emission showed signs of absorption beyond what is expected for the measured $E(B-V)$, which also points to the nova being embedded in some pre-existing cloud.  The fast optical decline is also consistent with an embedded nova.  However, the long-term behavior of the optical and near-infrared light curves were unlike other novae known to have red giant companions.  We also did not observe the rapid decrease in N$_{\rm H}$ that previous X-ray observations of novae with known red giant companions have reported.  We produced a model of the quiescent system as a K giant with an accretion disk which is in good agreement with the observed optical and infrared magnitudes.  We therefore find that the companion star in this system is producing a significant stellar wind and is most likely a K giant, specifically a K3-5 III, which is in agreement with Munari et al. (2017).  Several symbiotic binary systems with K giant stars are known, and they are often referred to as ``yellow symbiotics" (e.g., Baella et al. 2016).

\acknowledgements
We thank the anonymous referee for their thoughtful criticism of this manuscript which led to significant improvements.  We are grateful to C.~C. Cheung for his insight into the operations of the \textit{Fermi Gamma-ray Space Telescope}.  We also appreciate the useful conversations about novae, shocks, and accretion at the Stellar Remnants at the Junction meeting and the Conference on Shocks and Particle Acceleration in Novae and Supernovae.
We thank the NRAO for the generous allocation of VLA and VLBA time for our observations.  We also thank Neil Gehrels and the \textit{Swift} mission for the generous allocation of target-of-opportunity time to observe V1535 Sco.
We acknowledge support from NASA award NNX14AQ36G.
T.~N. was supported in part by NASA award NNX13A091G.
J.~L.~S. and J.~H.~S.~W. were funded in part by NSF award AST-1211778.
The National Radio Astronomy Observatory is a facility of the National Science Foundation operated under cooperative agreement by Associated Universities, Inc. Access to SMARTS has been made possible by generous support from the office of the Provost of Stony Brook University. This research made use of \emph{SciPy}, an open-source library of numerical routines for scientific computing available at http://www.scipy.org; \emph{APLpy}, an open-source plotting package for Python hosted at http://aplpy.github.com; and \emph{Astropy}, a community-developed core Python package for Astronomy (AstroPy Collaboration, 2013) available at http://www.astropy.org.  Many of the figures in this manuscript were made using the matplotlib 2D graphics package for Python.\\

\facility{Karl G. Jansky VLA, VLBA, Swift, SMARTS}

\software{AIPS, difmap, CASA, XSpec, SciPy, APLpy, AstroPy, matplotlib, IDL}

\end{document}